\documentclass[namedreferences]{solarphysics}

\usepackage[hyperref,optionalrh,showbiblabels]{spr-sola-addons} 
\usepackage{graphicx}        
\usepackage{color}           
\usepackage{breakurl}        

\chardef\us=`\_

\begin{document}

\begin{article}

\begin{opening}

\title{Coronal Hole and Solar Global Magnetic Field Evolution in 1976\,--\,2012}

%
\author{Irina~\surname{A.~Bilenko}$^{1}$\sep
       Ksenia~\surname{S.~Tavastsherna}$^{2}$\sep
       }

%
 \runningauthor{I.A.~Bilenko, K.S.~Tavastsherna}
 \runningtitle{Coronal Hole and Solar Global Magnetic Field Evolution in 1976\,--\,2012 }

%
  \institute{$^{1}$ Moscow M.V. Lomonosov State University, Sternberg Astronomical
  Institute, Universitetsky pr.13, Moscow, 119992, Russia,
                     email: \url{bilenko@sai.msu.ru}    \\
           $^{2}$ Central (Pulkovo) Astronomical Observatory, Russian Academy of Sciences,
                     Pulkovskoe sh. 65, St. Petersburg, 196140, Russia\\
           }

\begin{abstract}
Coronal hole spatial-temporal evolution is studied and comparison made with that
of the solar global magnetic field in cycles 21\,--\,23 (1976\,--\,2012).
The latitude-longitude distribution dynamics of coronal holes and the regularities
in the global magnetic field associated with the solar polar field reversal are analyzed.
Polar and non-polar coronal hole populations are considered.
The investigation reveals some temporal and spatial regularities in coronal hole
distributions that match well the global magnetic-field cycle evolution.
The results show that the non-polar coronal hole longitudinal distribution
follows all configuration changes in the global magnetic-field structure.
Reorganizations of the global magnetic-field and coronal hole distributions
occur simultaneously during a time interval of a few solar rotations.
The cycle evolution of the non-polar coronal holes
reflects the transition of the solar global magnetic field from the zonal structure to
sectorial and {\it vice versa}.
Two different type waves of non-polar coronal holes are revealed from their latitudinal distribution.
The first one is short poleward waves. They trace the poleward motion of the unipolar photospheric magnetic
fields from approximately $35^{\circ}$ to the associated pole in each hemisphere
and the redevelopment of a new-polarity polar CH. Although they start the poleward movement
before the change of the polar magnetic field in the associated hemisphere,
they reach the pole after the polar reversal.
The other type of non-polar CH wave forms two sinusoidal branches associated with the positive- and negative-polarity magnetic fields. The complete period of the wave was equal to $\approx$268 CRs (22 years).
These wave CHs arrive at high latitudes during declining phases when the new polarity polar CHs are already completely formed.
\end{abstract}

%

\end{opening}

%
 \section{Introduction}  \label{intro}

Coronal holes (CHs) are regions of low radiation in the extreme ultraviolet and X-ray
wavelengths in the solar atmosphere.
The study of CHs is important in view of their role in the
space weather formation at the Earth's orbit and their influence on geomagnetic activity.
CHs are associated with open magnetic field structures in the solar atmosphere and
they are considered to be the source of high-speed solar wind
(\opencite{Nolte1976}; \opencite{Obridko2011}; \opencite{Bilenko2005}; \opencite{Tlatov2014}).
CHs are mainly located on unipolar magnetic fields with one predominant
polarity (\opencite{Bohlin1978}; \opencite{Bumba1995}; \opencite{Harvey1982};
\opencite{Timothy1975}; \opencite{Varsik1999}).
\cite{Tavastsherna2014} found that 70\% of CHs were located in unipolar regions.
The unipolarity degree of the photospheric magnetic field in CH regions is 0.1\,--\,0.3 (\opencite{Tavastsherna2004}).
CHs are associated with low photospheric magnetic fields $\approx$1\,--\,5 G,
but they are located at the hills of the coronal magnetic field (\opencite{Obridko1989};
\opencite{Tavastsherna2004}).
Approximately 85\% of CHs are entirely or partly located in the regions of maximum coronal field
intensity for a given rotation (\opencite{Obridko1989}).
The average size of the photospheric magnetic elements of dominant polarity
and their parameters such as
magnetic-field strength, magnetic flux, and magnetic flux imbalance
in CH regions differ from that in ``quiet'' regions and the differences increase with solar activity cycle
(\opencite{Belenko2001}).
Since CHs are located in regions with a pronounced dominance of one of
the polarities of the solar magnetic field, the changes in their distribution over the
solar disk can be used as good tracers of evolutionary changes in the associated
positive- and negative-polarity magnetic field.

As shown by \cite{Ivanov2014}, the large-scale structure of the solar magnetic field determines the global organization of almost all solar activity phenomena, such as active regions, filaments, CHs, and coronal mass ejections. It was found that the time-space distribution of CHs on the solar disk is not uniform.
They form some cluster structures. The complexity of the structures and the
life-time of individual clusters depend on the solar cycle phase (\opencite{Bilenko2004b}).
It was also found that the longitudinal distribution of positive- and negative-polarity CHs
match well to that of the solar global magnetic field (GMF) in Cycle 23 (\opencite{Bilenko2012}).
\cite{Wang1990} noted that the topology of CHs is determined by that of the
unipolar regions in which they are embedded.

CHs can be divided into two groups: polar and non-polar CHs
according to their latitudinal location on the solar disk (\opencite{Sanchez1992}).
Their number, latitude distribution, rotation, and cycle evolution are different (\opencite{Insley1995}).
Polar CHs have a maximal area at the minimum phase of a
solar cycle (\opencite{Bravo1997}; \opencite{Dorotovic1996}; \opencite{Harvey2002}; \opencite{Hess2014}).
During the rising phase
polar CHs shrink and  they disappear completely about one to two years before the
sunspot maximum (\opencite{Waldmeier1981}).
At the maximum phase, the polar magnetic fields change their polarities
and new-polarity polar CHs are created at the solar poles
(\opencite{Webb1984}; \opencite{Fox1998}; \opencite{Harvey2002}, \opencite{Bilenko2002}).
The total number and area of non-polar CHs increase with the solar cycle progression
from the minimum to the maximum phase
(\opencite{McIntosh1992}; \opencite{Belenko2001}; \citeyear{Bilenko2002}).
It was found that the long-lived non-polar CHs have approximately the
same differential rotation as sunspots, and long-lived polar CHs show a
rigid rotation (\opencite{Ikhsanov1999}).
Polar and low-latitude CHs have different plasma parameters
(\opencite{Miralles2001}, \citeyear{Miralles2002}, \citeyear{Miralles2006}).
Polar CH cycle evolution is closely connected to the polar magnetic fields,
which confirms that global processes are involved (\opencite{Fox1998}).
\cite{Ikhsanov2015} showed that large CHs associated with high-latitude
large-scale magnetic fields and low-latitude small CHs that appear as a result
of decaying local sunspot fields formed two magnetic field systems, that
evolve in antiphase with respect to one another, with a shift in their minima by
$\approx$5\,--\,6 years.

Non-polar and polar CHs can also be divided into two subclasses, based
on CH area behavior at various phases of a solar cycle, on their latitude
and longitude distribution, rotation, and life-time as shown by \cite{Ikhsanov1999}.
They also found that at the rising and maxima phases, the CH magnetic fields
are of quadrupole type and at the declining and minima phases they are of dipole type.
The period of quadrupole magnetic field coincides with the epoch of more rigid
equatorial CH rotation because of high-latitude CH emergence.
It was shown that the low-latitude long-lived CHs possess differential rotation
that is similar to that of sunspot groups, and the long-lived polar CHs revealed a rigid rotation (\opencite{Ikhsanov1999}).

Occurrence and cycle evolution of CHs depend also on their association with active regions
or GMF (\opencite{Bilenko2004a}). Active-region CHs are closely connected to the processes
occurring in the active regions. CHs which are not connected to active regions reflect the
distribution of large-scale solar magnetic fields determined by the
GMF cycle evolution from zonal structure at the solar minimum to sectorial at the maximum when an alternation of longitudes covered by opposite-polarity CHs is observed (\opencite{Bilenko2002}).
The power spectra of CH numbers have periods of 13, 27, and 160 days.
These are the same periods which are detected for other solar
activity phenomena. They reflect the two and four sectorial structures of the GMF  (\opencite{Belenko2001}).
\cite{Bumba1995} found that  two opposite longitudinal rows of equatorial CHs located approximately 180$^{\circ}$ apart existed in 1991. They concluded that the formation of CHs at each such active longitude was a global process, and that this process depends on the development and distribution of background and local magnetic fields and on the phase of the cycle. The nine-day period recurring for several solar rotations, revealed in solar wind parameters, is also the manifestation of the periodic longitudinal distribution of CHs on the Sun (\opencite{Temmer2007}).

The aim of this work is to analyze the spatial and temporal distributions of CHs
on the solar disk, their relation to the solar polar magnetic-field reversal
and to compare them to that of the GMF during 1976\,--\,2012.
An understanding of the observed CH and GMF cycle spatial-time evolution should
provide some insight into the nature of solar cycle formation and dynamics.

Section \ref{data} describes the data used.
In Section \ref{secgar}, the spherical harmonic description
of the solar GMF is presented.
CH latitudinal distribution in cycles 21\,--\,23 is analyzed in Section \ref{secchlat}.
The longitudinal distributions of CHs and the GMF are investigated in Section \ref{secchlon}
The comparison of CH parameters and GMF is made in Section \ref{secchparam}.
The main conclusions are listed in Section \ref{secconclusion}.

\section{Data}   \label{data}

The CH catalog of the Solar Kislovodsk Mountain Astronomical Station of
Pulkovo Observatory was used. The catalog includes data on CH locations:
latitude and longitude and CH parameters
such as CH extension in latitude and longitude, area, associated photospheric
magnetic-field strength (in G) and polarity, and magnetic flux. The detailed
description of the method used for CH definition, CH parameter determination,
and the catalog creation are described in \cite{Tlatov2014}.

To compare the solar GMF and CH evolution at different phases of
solar cycles 21\,--\,23, we use data on the daily photospheric large-scale magnetic fields
and calculated coronal magnetic fields
from the Wilcox Solar Observatory (WSO) for the years 1976\,--\,2012.
Daily photospheric magnetograms are full-disk maps of the line-of-sight component of magnetic flux at the photosphere. The line used is the 5250 \AA  \, absorption line of neutral iron (Fe I).
The coronal magnetic field is calculated from
photospheric fields with a potential field model with the
source-surface location at 2.5 {\it R}$\odot$ (\opencite{Schatten1969};
\opencite{Altschuler1969}, \citeyear{Altschuler1975};
\opencite{Hoeksema1984}; \opencite{Hoeksema1986}, \citeyear{Hoeksema1988}).
Source-surface magnetic-field data consist of $30$ data points in equal steps
of sine latitude from $+70^{\circ}$ to $-70^{\circ}$. Longitude is
presented in $5^{\circ}$ intervals.
Full-disk synoptic map data span a full Carrington Rotation (1 CR = 27.2753 days).
The entire data set consists of 488 synoptic maps and covers CRs
1642\,--\,2130 (June 1976\,--\,November 2012).

Daily coronal hole maps in
the $\lambda = 10 830$ \AA \, line of the Kitt Peak NSO, SOHO/EIT (Solar and Heliospheric Observatory/Extreme ultraviolet Imaging Telescope) (\opencite{Delaboudiniere1995}) images in the $\lambda = 284$ \AA \, line, and {\it Yohkoh}/SXT (Soft X-ray Telescope) (\opencite{Tsuneta1991}) images were used for illustrations.

\section{Spherical Harmonic Description of the Solar Global Magnetic Field}   \label{secgar}

\begin{figure}
   \centerline{\includegraphics[width=1.\textwidth,clip=]{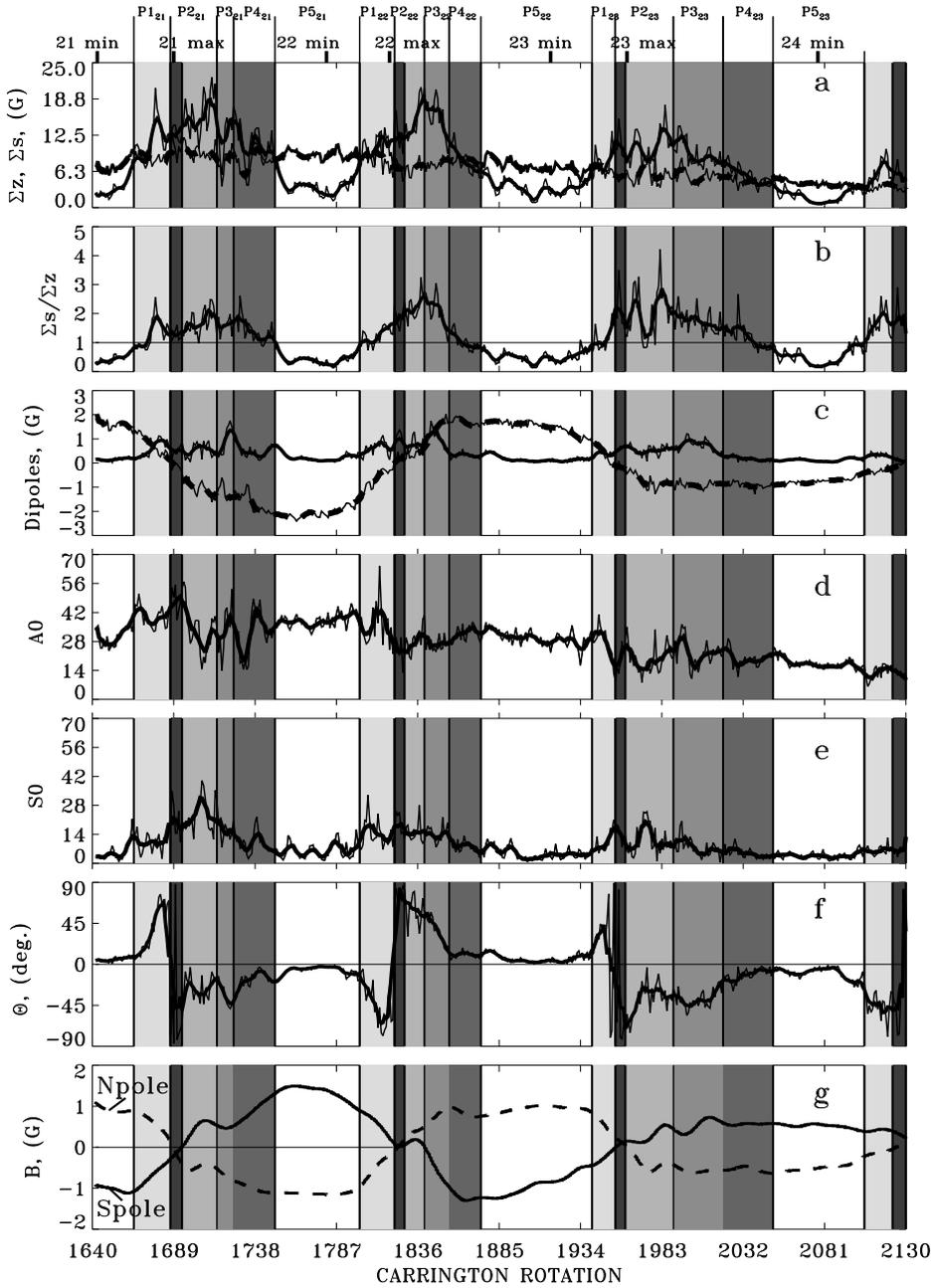}}
   \caption{(a) Sectorial harmonic spectra sum (solid line) and
                     that of zonal harmonics (dashed line);
               (b) the ratio of the sectorial harmonic sum to the zonal harmonic sum;
               (c) dipole components: axisymmetric dipole (dashed line) and equatorial dipole (solid line);
               (d) the axisymmetric, and with respect to the equator, antisymmetric harmonic spectra sum;
               (e) the axisymmetric, but symmetric with respect to the equator, harmonic spectra sum;
               (f) the polar angle, $\theta$, of the dipole component;
               (g) polar magnetic-field evolution in the northern and southern hemispheres.
                   Thin lines correspond to CR-averaged and thick lines represent seven CR-averaged data.
                   Light-, mid-, and dark-grey mark the P1\,--\,P4 periods of the sectorial GMF structure domination.
                   Black indicates a magnetic-field polarity reversals at the North and South poles.}
   \label{gar}
 \end{figure}

We use spherical harmonic analysis to investigate the solar GMF
cycle evolution during 1976\,--\,2012. The magnetic field can be described as a function of latitude
and longitude coordinates ($r, \theta, \phi$)  by the potential function
(\opencite{Altschuler1969}; \opencite{Altschuler1975}; \opencite{Altschuler1977};
\opencite{Chapman1940}):

$$ \psi(r,\theta,\phi)=R \sum_{n=1}^N \sum_{m=0}^n
\left(\frac{R}{r}\right)^{n+1} [g_n^m \, {\rm cos}\left(m\phi\right)\, + \, h_n^m \, {\rm sin}\left(m\phi\right)] \, P_n^m(\theta),  \label{psi}
$$

\noindent where  $P_{n}^m(\theta)$ are the associated Legendre
polynomials, and $N$ is the number of harmonics. The coefficients
$g_n^m$, $h_n^m$ are calculated using a least mean-square fit to
the observed line-of-sight photospheric magnetic fields with a
potential field assumption. The harmonic power spectra can be
calculated (\opencite{Altschuler1977}; \opencite{Levine1977}) using

$$
 S_n=\sum_{m=0}^n [(g_n^m)^2 + (h_n^m)^2].
$$

The sectorial harmonic spectra sum (solid line) and that of zonal harmonics
(dashed line) are presented Figure~\ref{gar}a.
In cycle 23, the sectorial components have a long ``tail''.
The ratio of the sectorial to zonal harmonic sum is presented in
Figure~\ref{gar}b. The horizontal line marks the level where the sum
of the sectorial harmonics is equal to that of the zonal harmonics.

The temporal evolution of the axisymmetric harmonic component of the solar dipole $g_1^0$ (dashed line)
and that of the equatorial dipole (solid line)

$$
S_{eqv.dip.} =   \sqrt{(g_1^1)^2 + (h_1^1)^2}
$$

\noindent are shown in Figure~\ref{gar}c.

In Figure~\ref{gar}d, the axisymmetric, and with respect to the equator, antisymmetric
harmonic spectral sum (\opencite{Stix1977}) is shown:

$$
A0=\sum_{n=1,3,5}(n+1)g_n^0 P_n(\theta).
$$

In Figure~\ref{gar}e, the axisymmetric, but symmetric with respect to the equator,
harmonic spectra sum (\opencite{Stix1977}) is presented:

$$
S0=\sum_{n=2,4,6}(n+1)g_n^0  P_n(\theta).
$$

The polar angle, $\theta$, of the dipole component evolution is shown in Figure~\ref{gar}f:

 $$
    \tan\theta=(g_1^0)^{-1} [(g_1^1)^2 + (h_1^1)^2]^{1/2}.
 $$

The observed North pole (dashed line) and South pole (solid line) magnetic-field
variations are shown in Figure~\ref{gar}g.
The maxima and the minima of cycles 21\,--\,24 are marked at the top of Figure~\ref{gar}.

For further investigation and comparison of the GMF and CHs,
different periods are selected in each cycle according to the GMF harmonic spectra
and GMF structure (Figures~\ref{gar}, \ref{lon}).
The first periods (P$1_{21}$\,--\,P$1_{23}$) are the time from the beginning of the
sectorial structure domination to the beginning of the polar
magnetic-field sign changes in each cycle.
During the first period, the polar magnetic-field strength,
zonal harmonic spectra sum, and axisymmetric  component of the solar dipole decreased.
The sectorial components increased.
The non-axisymmetric component of the GMF, A0, was at the highest level in
each cycle. The axisymmetric component of the GMF, S0, increased,
and the inclination of $\theta$ increased rapidly.

Periods P$2_{21}$\,--\,P$2_{23}$ and P$3_{21}$\,--\,P$3_{23}$ are defined as being from
the beginning of the polar magnetic-field sign changes
to the end of two-sectorial structure domination (Figure~\ref{lon}),
when the sum of sectorial components becomes approximately
equal to that of zonal components, and $\theta$ reaches a minimal value.
During these periods the sectorial structure dominates.
The axisymmetric  component of the solar dipole changes sign and
begins to increase in each cycle.
The non-axisymmetric component of the GMF (A0) reaches a minimum,
and the axisymmetric component of the GMF (S0) increases to a maximum.
The magnetic field at the poles changes sign and begins to increase.
The polar angle, $\theta$, reaches the highest latitudes and begins to decline.

The periods (P$4_{21}$\,--\,P$4_{23}$) last until the end
of the sectorial structure domination in each cycle. During these periods
the magnitude of sectorial components diminishes and
that of the zonal components increases.
The axisymmetric component of the dipole and
the polar magnetic-field strength increase.
The non-axisymmetric component of the GMF (A0) increases slightly,
and the axisymmetric component of the GMF (S0) decreases.
The polar angle, $\theta$, is reduced to zero.

The periods (P$5_{21}$\,--\,P$5_{23}$) are characterized by
the zonal GMF structure domination.
The polar magnetic-field strength and the axisymmetric component
of the solar dipole reach a maximmum.
The axisymmetric component of the GMF (S0) is at the lowest level.
The polar angle, $\theta$, is close to zero.
In Figure~\ref{gar}, the periods are marked in light-, mid-, and dark-grey.
Black indicates magnetic-field-polarity sign changes at the North and South poles.

The sectorial and zonal harmonic spectral sum, the magnitude of axisymmetric
component of the solar dipole, A0, S0, and the North and South polar magnetic-field
strength diminish from cycle 21 to cycle 23.

\section{Latitudinal Distribution of Coronal Holes}   \label{secchlat}

\begin{figure}
   \centerline{\includegraphics[width=1.\textwidth,clip=]{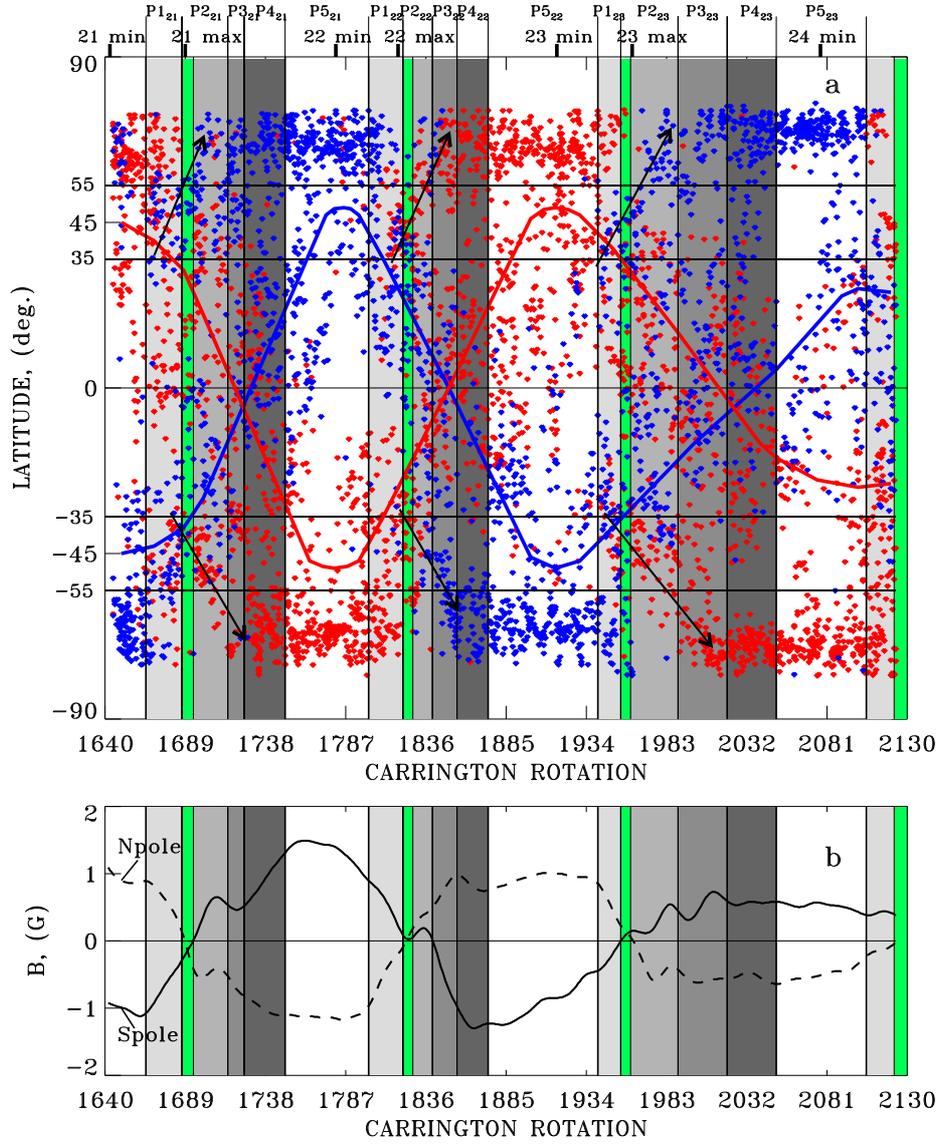}}
      \caption{(a) The latitudinal distribution of CHs in 1975\,--\,2012;
                  (b) polar magnetic-field evolution in the northern and southern hemispheres.
      Red denotes the CHs associated with the positive-polarity photospheric magnetic fields and
      Blue denotes the CHs associated with the negative-polarity photospheric magnetic fields.
      Light-, mid-, and dark-grey  mark the P1\,--\,P4 periods of the sectorial GMF structure domination.
      Green indicates magnetic-field polarity reversals at the North and South poles.}
   \label{lat}
 \end{figure}

Figure~\ref{lat} presents (a) the latitudinal distribution of positive-polarity (red)  and
negative-polarity (blue) CHs and (b) North and South polar magnetic fields in 1975\,--\,2012.
The maxima and minima of cycles 21\,--\,24 are marked at the top of Figure~\ref{lat}.

CHs are clearly divided into two groups: polar and non-polar CHs, according to their
latitudinal locations. Thick horizontal lines at $\pm55^{\circ}$ latitudes separate
the two groups.

The two different type waves of non-polar CHs can be detected from their latitudinal
distribution shown in Figure 2.
The first one (waves 1) are short poleward waves that are marked by black arrows.
They indicate new-polarity CHs traveling from approximately $35^{\circ}$ to the
associated pole in each hemisphere.
\cite{Ikhsanov2013} also found that the large-area ($>$15000 millionths of the solar
hemisphere) high-latitude CHs and polar faculae rose from midlatitudes ($40^{\circ} - 50^{\circ}$)
to the polar regions.The rise was in the form of several slaned chains about a half-year width in
time and with a periodicity of 1.25$\pm0.3$ years in cycle 21. They noted that
the CH polarity corresponded to the trailer polarity of active regions.

The other type are non-polar CH waves (waves 2) that form two sinusoidal branches.
One branch is associated with
the positive-polarity magnetic fields (red line) and the other with that of the negative-polarity (blue line).
Wave 2 CHs associated with the positive- and
negative-polarity magnetic fields are in antiphase in each cycle.
The latitude-location cycle changes of the wave 2 CHs coincide with that of the
axisymmetric component of the solar dipole ($g_1^0$), as shown in Figure~\ref{gar}c.
Wave 2 CH branches and  the axisymmetric component of the solar dipole
are the lowest in cycle 23, where they do not exceed latitudes $\pm35^{\circ}$.
The complete period of the wave 2 is equal to $\approx$268 CRs ($\approx$22 years).
Since CHs are located in solar atmospheric regions with a dominance of one of
the magnetic-field polarities, the changes in their distribution over the
solar disk is the manifestation of evolutionary cycle changes in the solar magnetic fields.
When two branches of positive- and negative-polarity magnetic fields, traced by wave 2 CHs, move below $\approx$$\pm35^{\circ}$ in latitude, the sectorial structure of the GMF is established.
Unipolar CH longitudes are formed, which are occupied by only positive or by only
negative-polarity CHs alternated approximately with a period of 13 days and
covering $\approx$$60-100^{\circ}$ of longitude (\opencite{Bilenko2002}).
When both wave 2 polarity branches are located higher than $\approx$$\pm35^{\circ}$ and reach
the highest latitudes in each hemisphere, the zonal GMF structure was observed.
The polarity of CH-associated magnetic fields matches the polarity of the polar regions
in the associated hemisphere at that time.

The CH evolution during a solar cycle is closely related
to solar polar magnetic-field reversals (\opencite{Fox1998};
\opencite{Webb1984}; \opencite{Harvey2002}).
\cite{Webb1984} studying the evolution of the polar magnetic field around
sunspot maximum and polar CH redevelopment in cycles 21 and 22 and found
that the process of polar field reversals and redevelopment of the polar CHs was
discontinuous and occurred  in two or three longitudinal bands, and an asymmetry
of the processes in opposite hemispheres was revealed. The polarity reversals in
the two hemispheres were offset between six months to one and a half years.
\cite{Harvey2002} found that new-polarity CHs
formed in the remnants of the follower active region magnetic fields before the
polar reversal and expanded to cover the poles within three solar rotations
after the polar reversal in cycles 22 and 23. During the first 1.2\,--\,1.4
years polar CHs  were asymmetric (\opencite{Harvey2002}).

From Figure~\ref{lat} it is seen that wave 2 CHs
are not associated with that process.
They arrive at high latitudes when the polar CHs of a new polarity have already completely formed.
Even more, they reach the highest latitudes and the locations of polar CH regions during the late decline phase.
The wave 2 CHs reach the highest latitudes during the period of the zonal
GMF structure domination (P$5_{21}$\,--\,P$5_{23}$).

As seen in Figure~\ref{lat}a, the wave 1 CHs form before
the polar magnetic-field reversal in each cycle.
The polarity of the wave 1 CHs corresponds to the trailer polarity
of active regions and, correspondingly, to the leading polarity of the next cycle (Figure~\ref{lat}a).
The waves 1 began $\approx$16 and $\approx$8 CRs before the polar
magnetic-field sign changes in the North and South pole correspondingly in cycle 21,
$\approx$10 and $\approx$3 CRs in cycle 22, and $\approx$20 and $\approx$18 CRs
in cycle 23. The first CHs with the polarity of the polar CH waves appear
around the time of the beginning of the sectorial stricture domination, the first period
(P$1_{21}$\,--\,P$1_{23}$), in each cycle.
The waves are asymmetric in the North and South hemispheres.
The longest waves 1 are seen in cycle 23.
Wave 1 CHs trace the poleward motion of the unipolar photospheric magnetic
fields to the polar regions and the redevelopment of a new-polarity polar CH.
Wave 1 CHs and magnetic flux start a poleward movement
before the change of the polar magnetic field in the associated hemisphere,
but they reach the pole after the reversal.
They reach the pole region and the main new-polarity polar CHs are formed at
the pole regions just after the polar magnetic-field
sign changes at the North pole in cycles 21, 22 and in $\approx$20 CRs after the polar field reversal
at the North pole in cycle 23, and approximately 40, 27, and 38 CRs
after the polar magnetic-field polarity changes at the South poles in
cycles 21, 22, and 23 correspondingly.
It should be noted, that some old-polarity polar CHs are observed at the time of the polar
field reversals and up to $\approx$20 CRs after the polar magnetic-field sign changes.
The episodical new-polarity CHs are observed $\approx$10\,--\,15 CRs before
the polar field reversal at the polar CH latitudes.

Polar CHs are arranged more compactly in cycle 23, and they are located in a
wider range of latitudes in cycles 21 and 22.

\section{Longitudinal Distribution of Coronal Holes and the Solar Global Magnetic Field}  \label{secchlon}

\begin{figure}
     \vspace{-0.2cm}
      \centerline{\includegraphics[width=1.\textwidth,clip=]{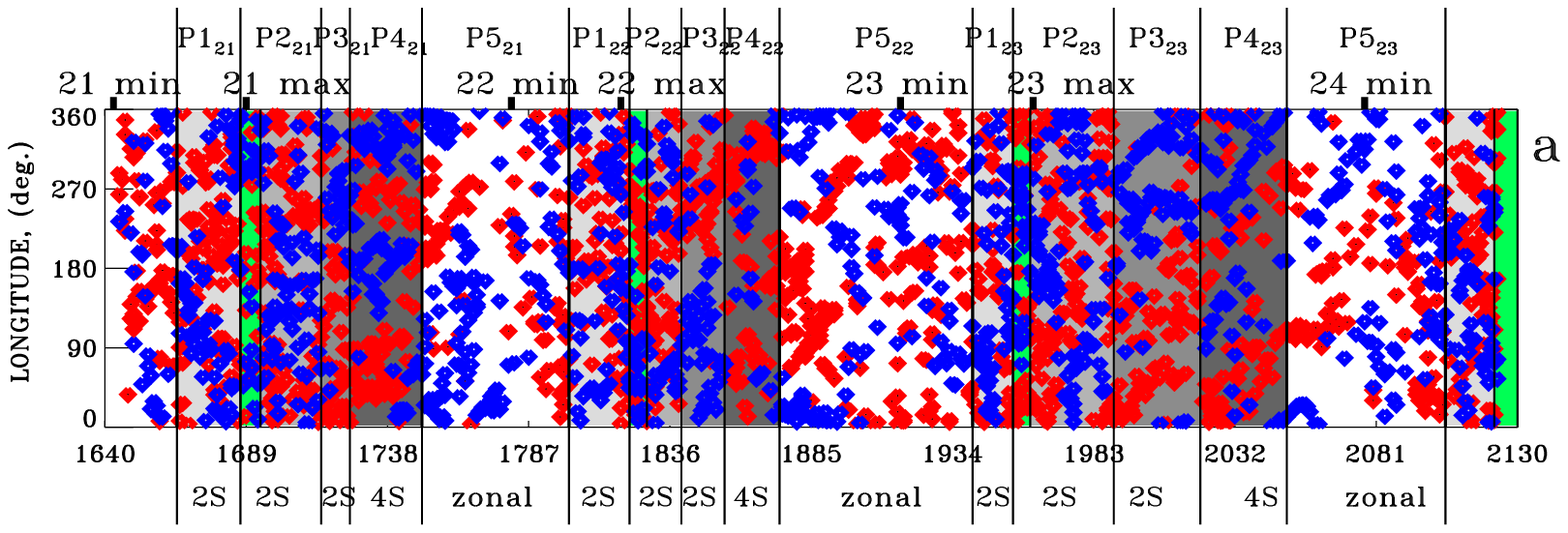}}
     \vspace{-0.2cm}
      \centerline{\includegraphics[width=1.\textwidth,clip=]{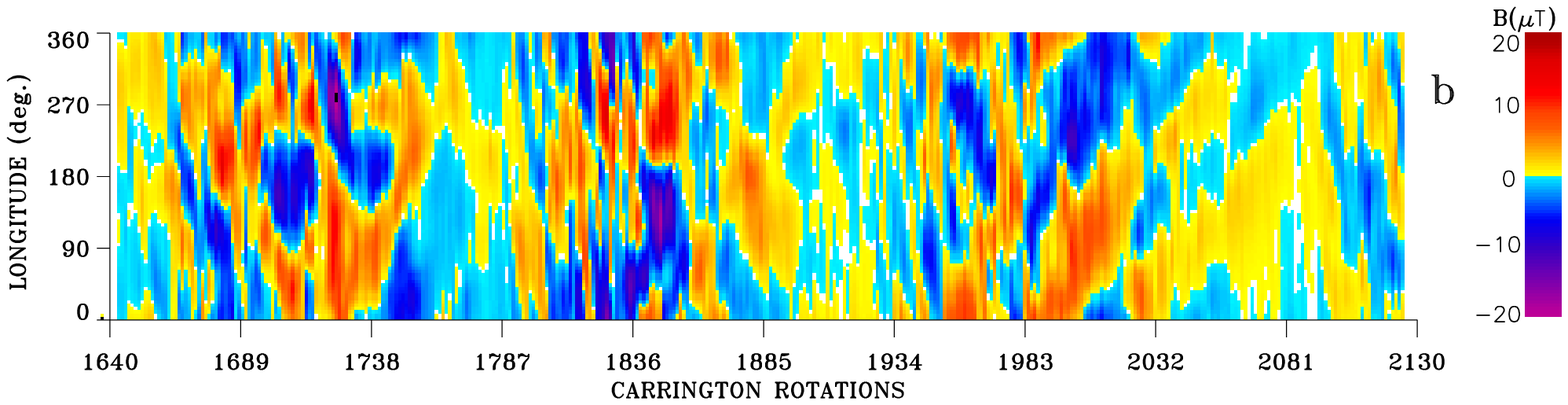}}
     \vspace{-0.65cm}
      \centerline{\includegraphics[width=1.\textwidth,clip=]{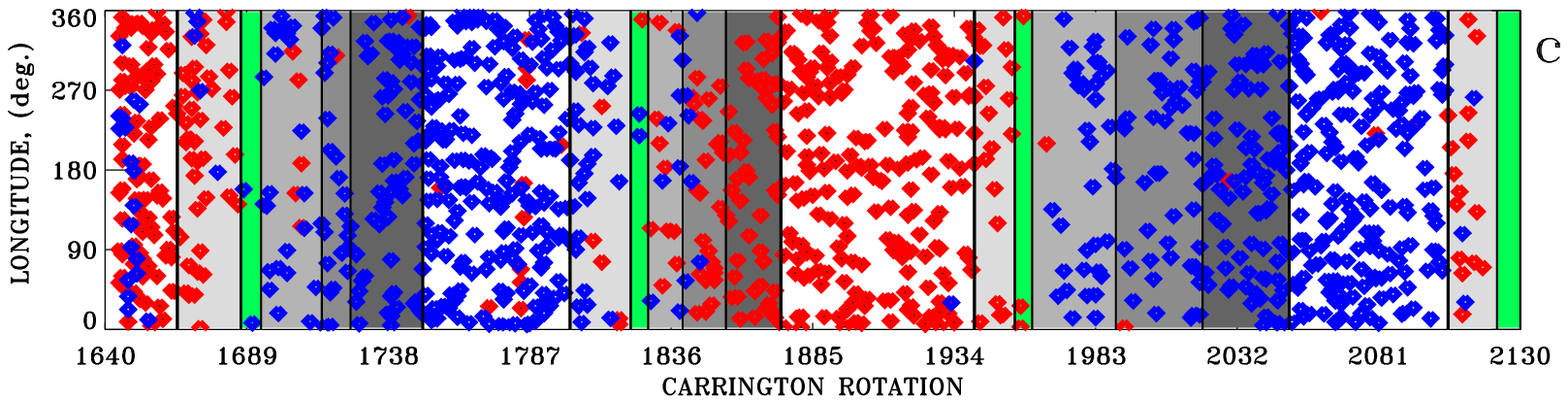}}
      \centerline{\includegraphics[width=1.\textwidth,clip=]{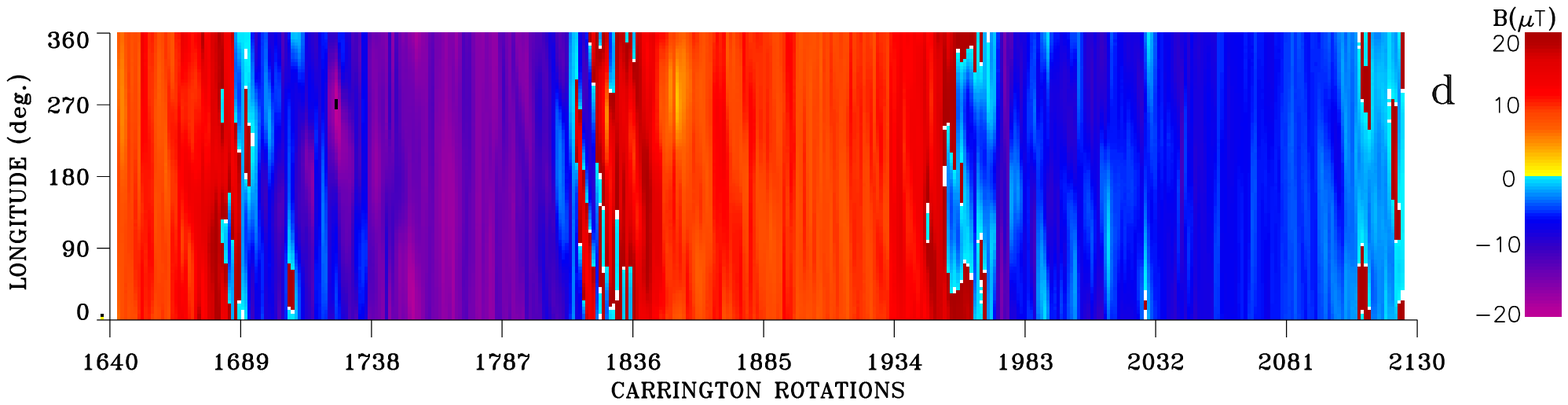}}
     \vspace{-0.65cm}
      \centerline{\includegraphics[width=1.\textwidth,clip=]{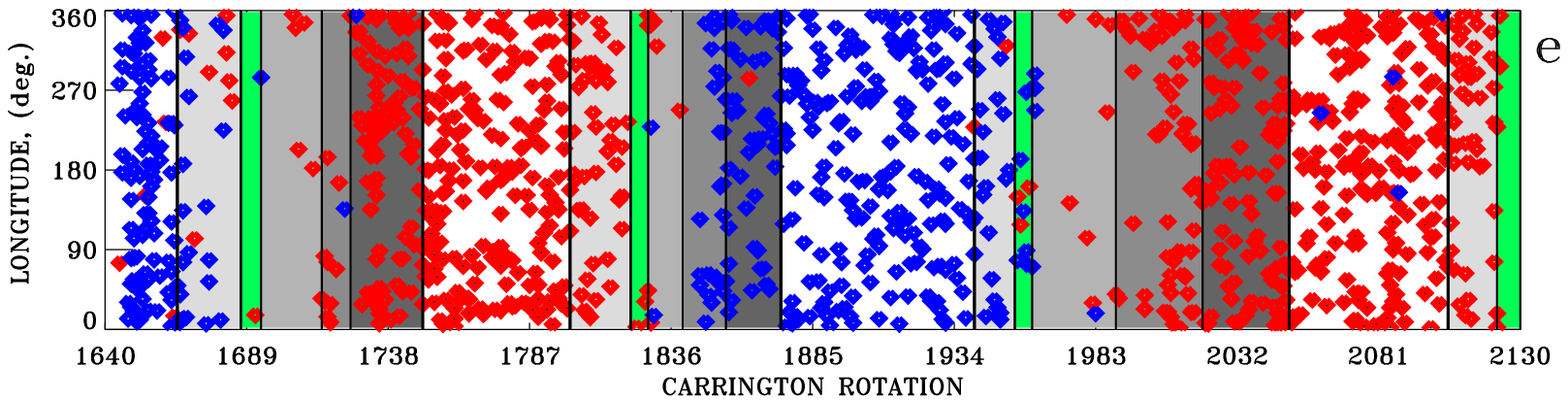}}
      \centerline{\includegraphics[width=1.\textwidth,clip=]{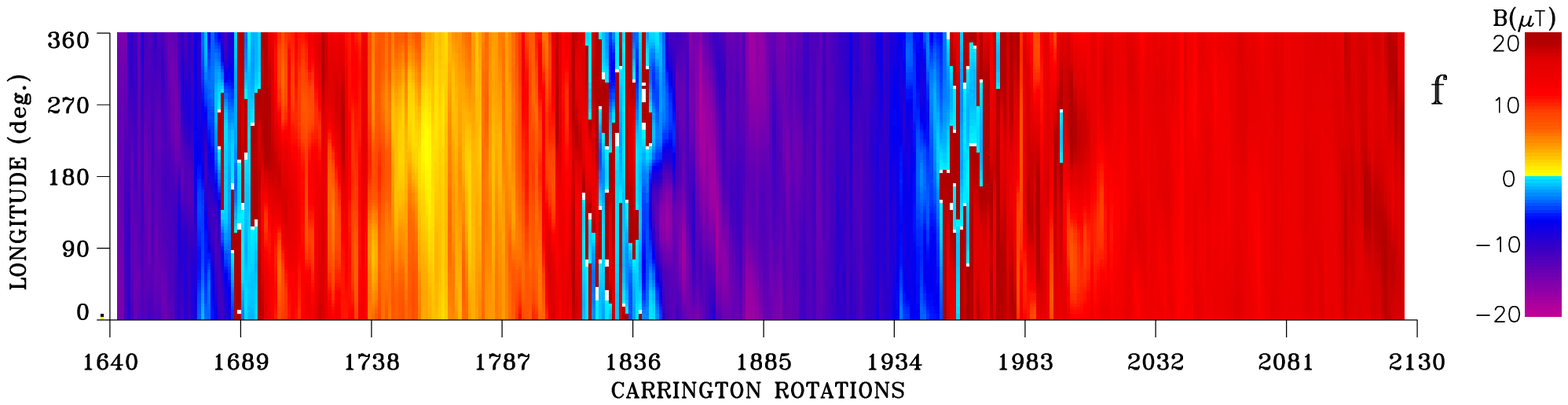}}
     \vspace{-0.2cm}
      \caption{Longitudinal distributions of: (a) non-polar CHs associated with the positive-polarity (red) and negative-polarity (blue) magnetic fields;
               (b) GMF from $-55^{\circ}$ to $+55^{\circ}$ latitude for positive-polarity (red) and
                    negative-polarity (blue) magnetic fields;
               (c) polar CHs associated with the positive-polarity (red) and negative-polarity (blue)
                   photospheric magnetic fields in the North hemisphere;
               (d) GMF from $+55^{\circ}$ to $+70^{\circ}$ latitude for positive-polarity (red) and
                    negative-polarity (blue) magnetic fields in the North hemisphere;
               (e) polar CHs associated with the positive-polarity (red) and negative-polarity (blue) photospheric magnetic fields in the South hemisphere;
               (f) GMF from $-55^{\circ}$ to $-70^{\circ}$ latitude for positive-polarity (red) and
                    negative-polarity (blue) magnetic fields in the South hemisphere.
               Light-, mid-, and dark-grey mark the P1\,--\,P4 periods of the sectorial GMF structure domination.
               Green indicates the magnetic-field polarity reversals at the North and South poles.}
   \label{lon}
\end{figure}

In Figure~\ref{lon}a, the longitudinal distribution of non-polar CHs is presented
and in Figure~\ref{lon}b, the time-longitude distribution
of the strength and polarity of GMF from $-55^{\circ}$ to $+55^{\circ}$ latitudes
is shown. The longitudinal diagram was created in a CR-rotation system.
The $x$-axis denotes the date of 0$^{\circ}$ CR longitude at the central
meridian, and the $y$-axis denotes longitude.
A detailed description of the longitudinal diagram creation and GMF changes
is given in \cite{Bilenko2014}. The maxima and minima of the cycles are
marked at the top of Figure~\ref{lon}.
The comparison of the CH and GMF longitudinal distributions shows that
the CH cluster structure (\opencite{Bilenko2004b}) completely coincides with the GMF
longitudinal distribution.
Positive-polarity CH locations coincide with positive-polarity (red) GMF and
negative-polarity CH locations coincide with negative-polarity (blue) GMF.
CH longitudinal distributions follow all configurations in the GMF.
Quasi-stable two-polarity GMF structures
match well to the positive- and negative-polarity CH locations at the solar maxima and at
the beginning of the declining phases.
Drifts in the GMF during the rising and declining phases
are also clearly revealed in CH longitudinal location changes.
During the late rise and maxima phases, the two-sector-polarity
GMF structures are formed (Figure~\ref{lon}b).
They are mostly occupied by positive- or by negative-polarity CHs (Figure~\ref{lon}a).
The two-polarity structure
in GMF and CHs is more pronounced during the declining phases in each cycle.
The periods of reorganization in the GMF,
characterized by the longitudinal rearrangement of magnetic-field polarity structures, are
repeated by the changes in the locations of CHs associated with the positive- and
negative-polarity photospheric magnetic fields. Reorganizations
of the GMF and CH cluster structures occur simultaneously
during a time interval of a few solar rotations. Old CH clusters disappear and a new set of CH
clusters forms following the reorganizations in the GMF (\opencite{Bilenko2004b}; \opencite{Bilenko2012}).

\begin{figure}
   \centerline{
             \includegraphics[width=1.\textwidth,clip=]{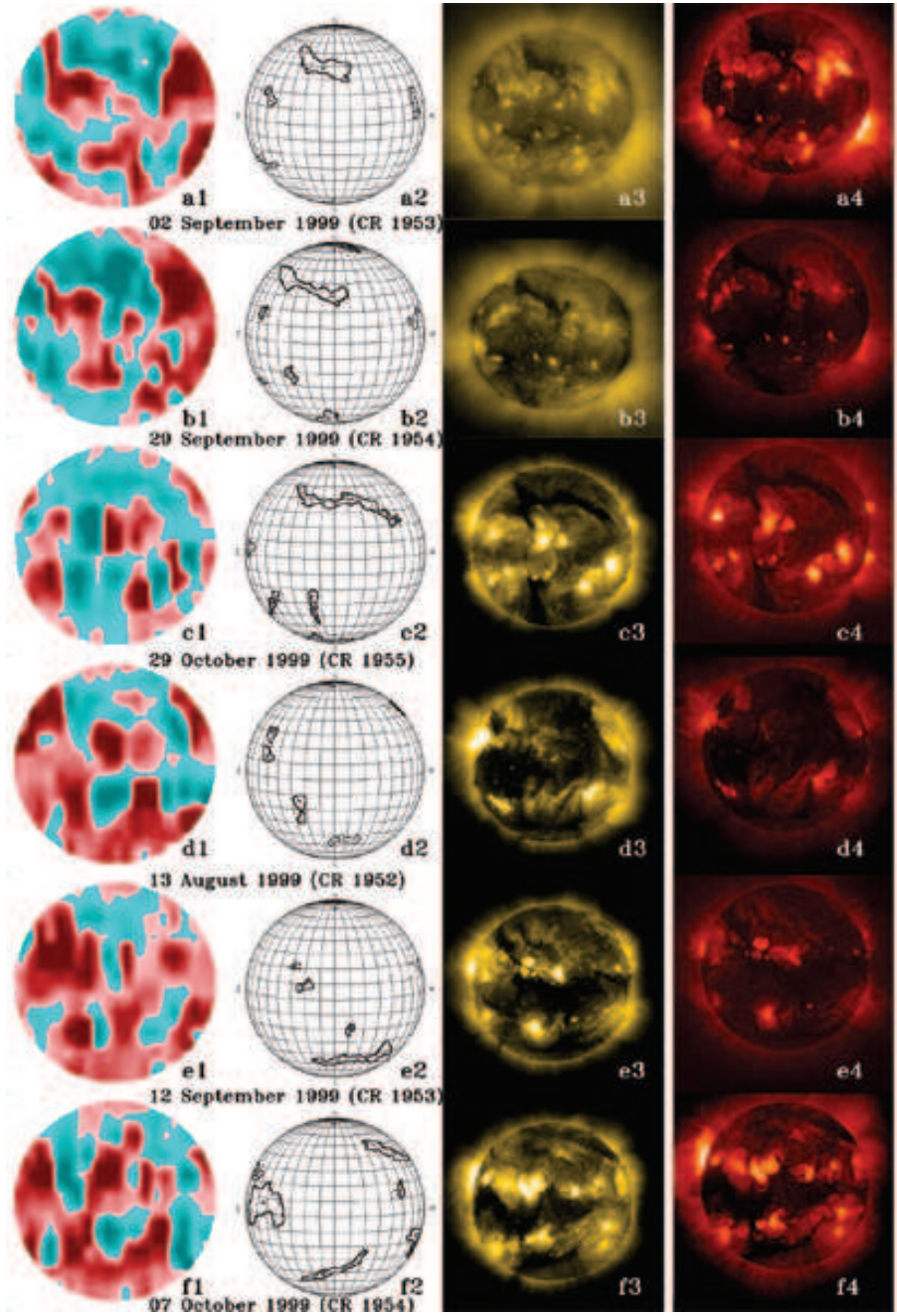}}
\caption{Large-scale photospheric positive-polarity (rose) and negative-polarity (blue) magnetic-field distributions (a1\,--\,f1) and wave 1 coronal holes in the $\lambda = 10 830$ \AA \, (a2\,--\,f2), $\lambda = 284$ \AA \, (a3\,--\,f3) lines, and soft X-ray (a4\,--\,f4) observed in the North (a\,--\,c) and South (d\,--\,f) hemispheres.}
   \label{wave1}
   \end{figure}

The drifts in the GMF and CHs are the manifestation of
the slower rotation at the rising phases and faster rotation at the declining phases.
\cite{Wang1990} proposed that rotation rate of CHs is determined by the centroidal latitude of the nonaxisymmetric flux. During the rising phase, CHs  rotate at the slow rate of the decaying
mid-latitude active-region magnetic fields. During the declining phase,
the equatorward extensions of the polar CHs rotate at the faster rate of
the low-latitude active-region remnants.
According to our results, from Figure~\ref{lat}a we can see that during
the rising phases, periods P$1_{21}$\,--\,P$1_{23}$,
old-polarity polar CHs and low-latitude CHs, and the associated photospheric magnetic fields,
are separated by the wave 1 CHs, that traced a new-polarity magnetic field.
As a result, they rotate with different rates.
In Figure~\ref{wave1}, the examples of daily large-scale photospheric magnetic-field distributions
and wave 1 coronal holes in the $\lambda = 10 830$ \AA, $\lambda = 284$ \AA \, lines, and soft X-ray
observed in the North (Figure~\ref{wave1}a\,--\,c) and the South (Figure~\ref{wave1}d\,--\,f) hemispheres are
presented. In the North hemisphere, negative-polarity (blue) magnetic-field structure separates
positive-polarity (rose) polar and low-latitude magnetic fields (Figure~\ref{wave1}a1, b1, c1).
The associated coronal hole is shown in EUV and X-ray  (Figure~\ref{wave1}a2\,--\,a4, b2\,--\,b4, c2\,--\,c4).
In the South hemisphere, positive-polarity magnetic-field structure separates negative-polarity south polar
and low-latitude magnetic fields (Figure~\ref{wave1}d1, e1, f1). The associated coronal hole is shown in EUV and X-ray  (Figure~\ref{wave1}d2\,--\,d4, e2\,--\,e4, f2\,--\,f4).
These coronal holes have an extended shape in longitude.

 \begin{figure}
   \centerline{
             \includegraphics[width=1.\textwidth,clip=]{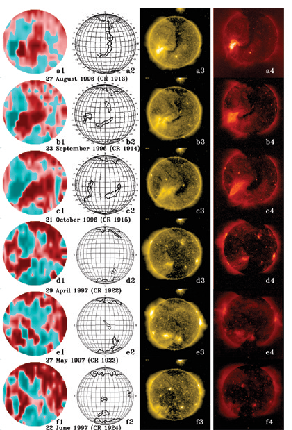}}
\caption{Large-scale photospheric positive-polarity (rose) and negative-polarity (blue)  magnetic-field distributions (a1\,--\,f1) and polar coronal-hole extensions in the $\lambda = 10 830$ \AA \, (a2\,--\,f2), $\lambda = 284$ \AA \, (a3\,--\,f3) lines, and soft X-ray (a4\,--\,f4) observed in the North (a\,--\,c) and South (d\,--\,f) hemispheres.}
   \label{ext}
   \end{figure}

During the declining and minimum phases, periods P$4_{21}$\,--\,P$4_{23}$, P$5_{21}$\,--\,P$5_{23}$,
the polarity of polar and non-polar CHs, and associated photospheric magnetic fields,
coincide. They create a common large-scale unipolar magnetic-field system.
Therefore, the slow rotation rate of the polar regions affects the deceleration
of the rotation of mid- and low-latitude photospheric magnetic fields
and, consequently, the associated CHs. Figure~\ref{ext} presents the examples of
daily large-scale photospheric magnetic-field distributions
and polar coronal-hole extensions in the $\lambda = 10 830$ \AA, $\lambda = 284$ \AA \, lines, and soft X-ray
observed in the North (Figure~\ref{ext}a\,--\,c) and the South (Figure~\ref{ext}d\,--\,f) hemispheres.
Positive-polarity magnetic-field structure is extended from the North pole to the South hemisphere high latitudes (Figure~\ref{ext}a1\,--\,c1). The associated coronal hole, the well known ``Elephant trunk'', is seen in EUV and X-ray (Figure~\ref{ext}a2\,--\,a4, b2\,--\,b4, c2\,--\,c4). Negative-polarity magnetic-field structure is extended from the South pole to the North hemisphere (Figure~\ref{ext}d1\,--\,f1). The associated coronal hole is shown in EUV and X-ray (Figure~\ref{ext}d2\,--\,d4, e2\,--\,e4, f2\,--\,f4). These coronal holes have an extended shape  in latitude. The polar CH extensions change their shape from CR to CR. The polar CH extensions to low latitudes rotate with the polar CH rate, even after they are disconnected (\opencite{Zhao1999}). The large-scale magnetic-field structures are more stable.

\cite{Fox1998} found that the polar CH evolution during solar activity cycles
was closely related to the polar magnetic fields.
Studying the polar magnetic-field
reversals in cycles 21 and 22, they suggested that
the polar reversals originated from the global processes rather than
from local magnetic flux dynamics.
In Figure~\ref{lon}c, the longitudinal distribution of the North polar CHs is presented
and in Figure~\ref{lon}d, the time-longitude distribution
of the strength and polarity of the GMF for latitudes from $55^{\circ}$ to $70^{\circ}$
is shown.
In Figure~\ref{lon}e, the longitudinal distribution of the South polar CHs is presented
and in Figure~\ref{lon}f the time-longitude distribution
of the strength and polarity of the GMF is shown for latitudes from $-55^{\circ}$ to $ -70^{\circ}$.

 \begin{figure}
 \vspace{-0.5cm}
   \centerline{
             \includegraphics[width=0.9\textwidth,clip=]{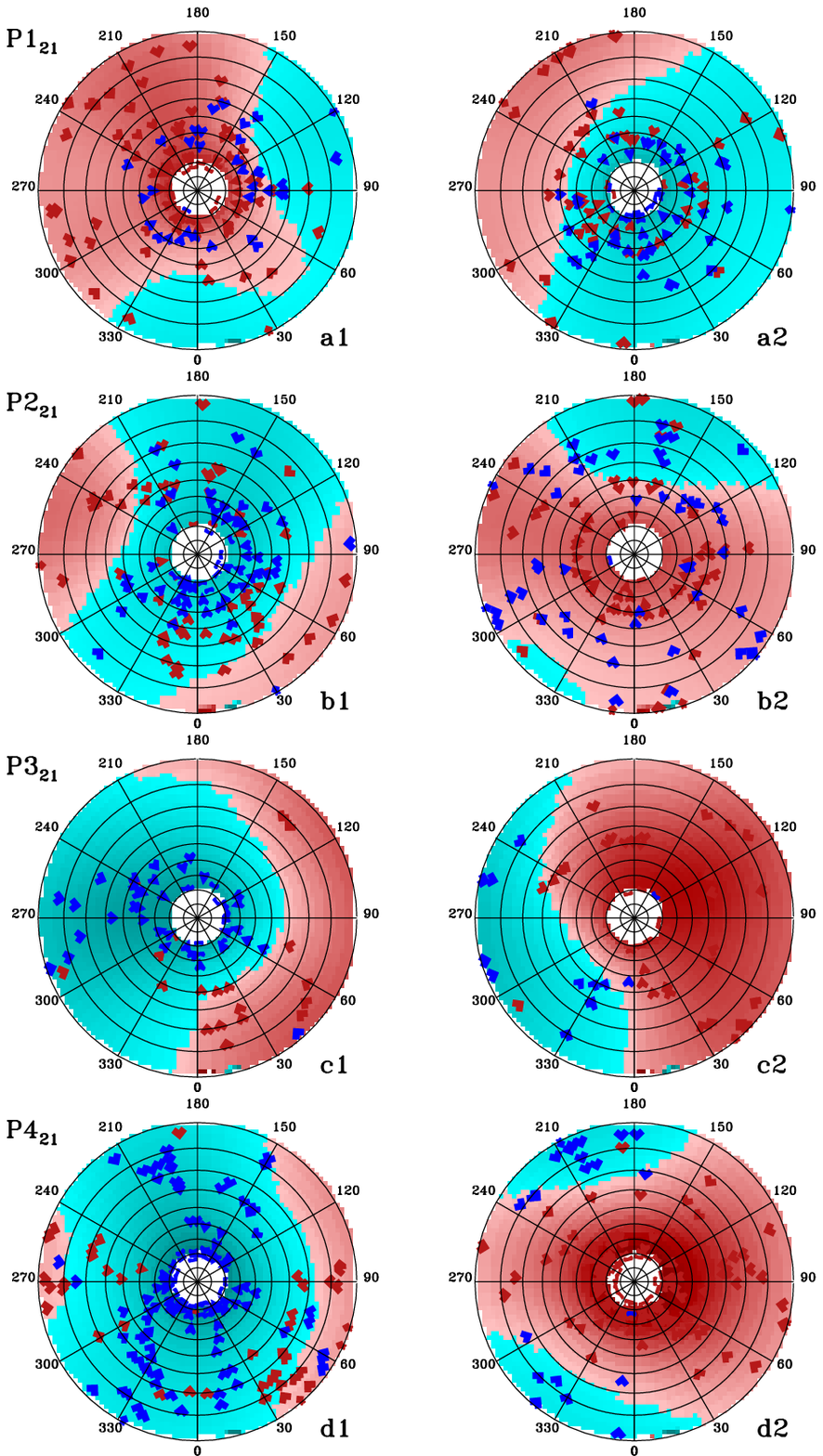}}
\caption{The polar projections of the distributions of CHs associated with positive-polarity (red) and negative-polarity (blue) photospheric magnetic fields and GMF seen from the North pole (left side panels) and South pole (right side panels) for periods P$1_{21}$\,--\,P$4_{21}$ in cycle 21.}
   \label{cycle21}
   \end{figure}

 \begin{figure}
   \centerline{
             \includegraphics[width=1.\textwidth,clip=]{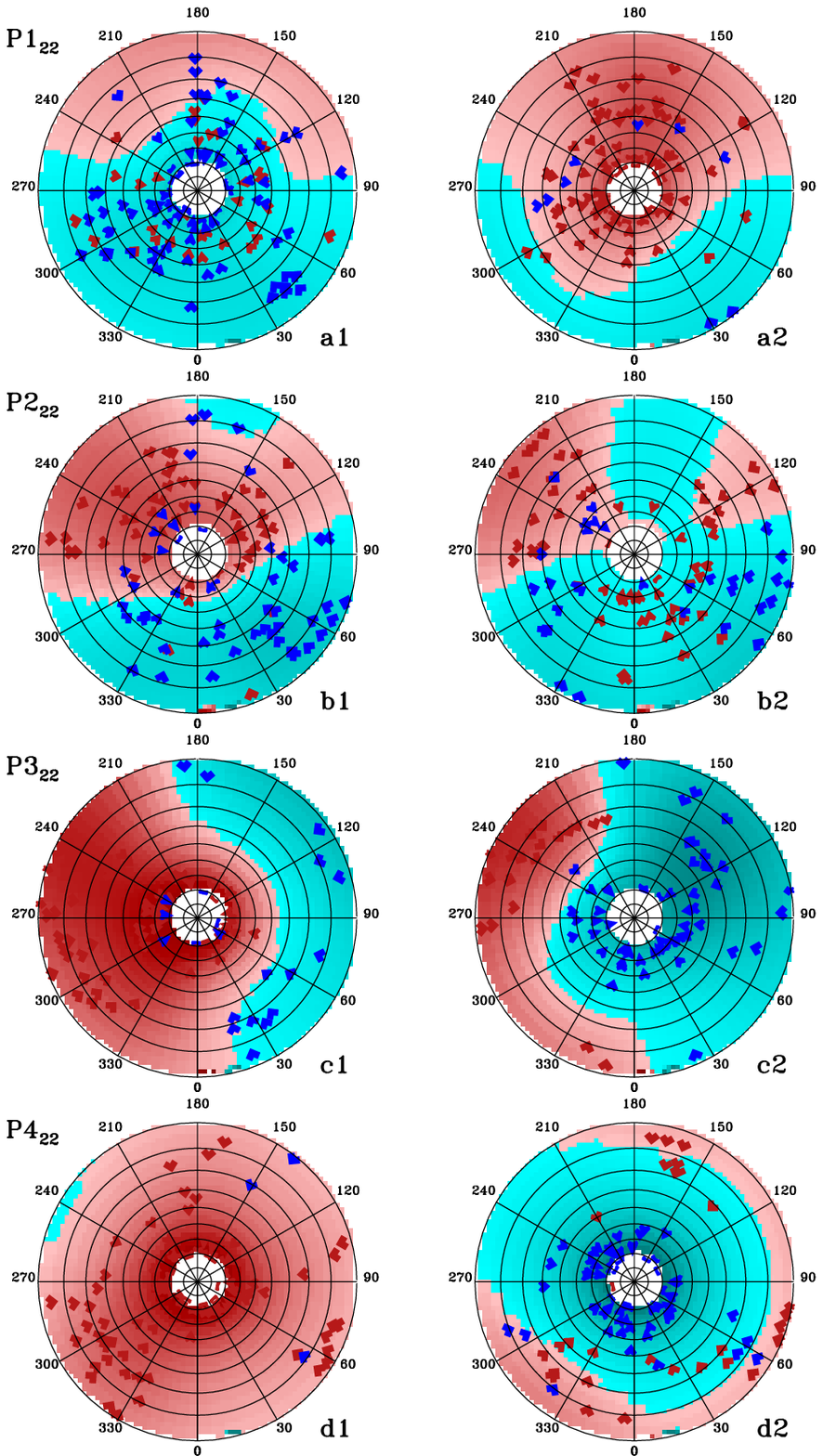}}
\caption{The polar projections of the distributions of CHs associated with positive-polarity (red) and negative-polarity (blue) photospheric magnetic fields and GMF seen from the North pole (left side panels) and South pole (right side panels) for periods P$1_{22}$\,--\,P$4_{22}$ in cycle 22.}
   \label{cycle22}
   \end{figure}

 \begin{figure}
   \centerline{
             \includegraphics[width=1.\textwidth,clip=]{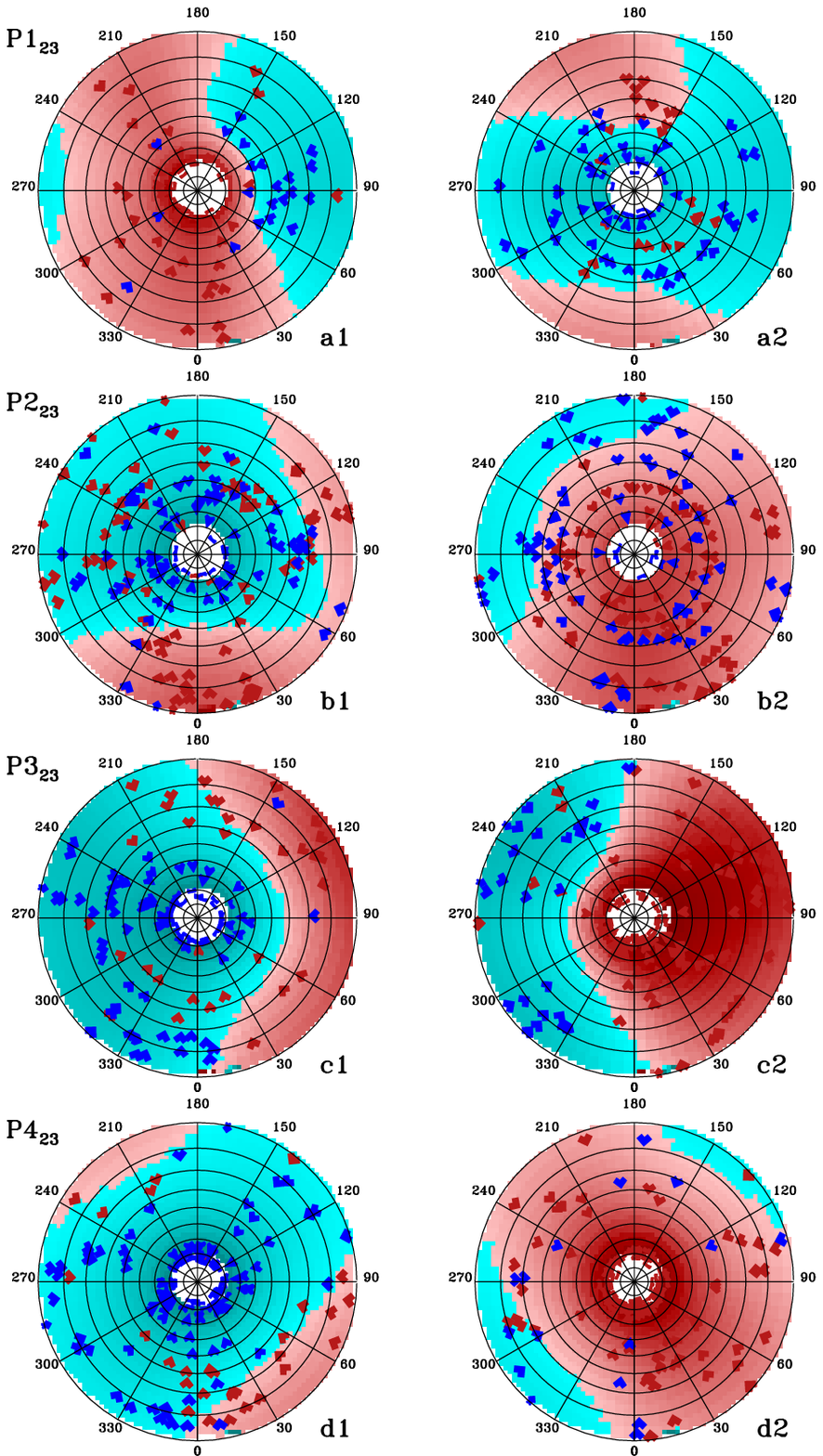}}
\caption{The polar projections of the distributions of CHs associated with positive-polarity (red) and negative-polarity (blue) photospheric magnetic fields and GMF seen from the North pole (left side panels) and South pole (right side panels) for periods P$1_{23}$\,--\,P$4_{23}$ in cycle 23.}
   \label{cycle23}
   \end{figure}

 \begin{figure}
 \vspace{-1.cm}
   \centerline{
             \includegraphics[width=1.\textwidth,clip=]{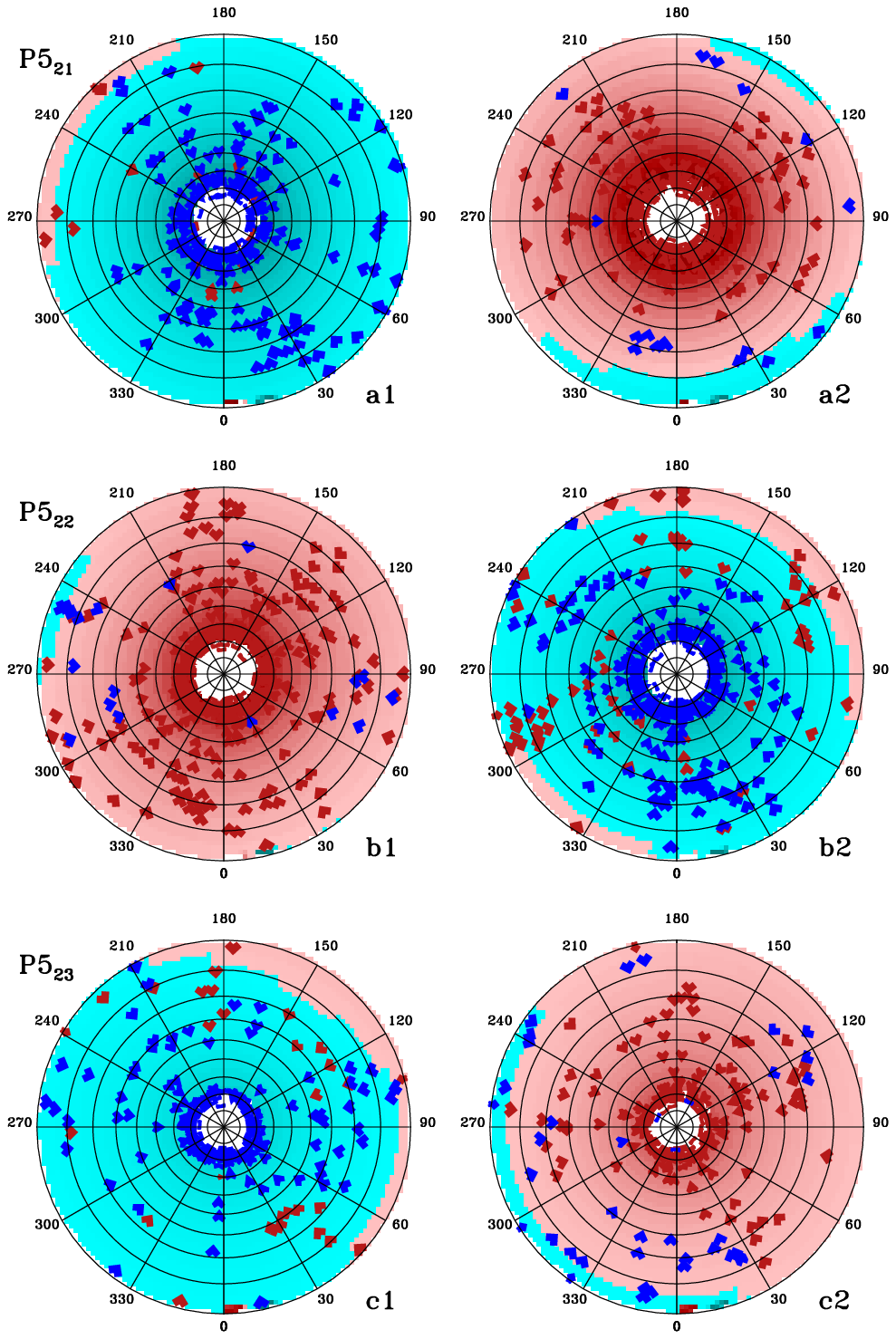}}
\caption{The polar projections of the distributions of CHs associated with positive-polarity (red) and negative-polarity (blue) photospheric magnetic fields and GMF seen from the North pole (left side panels) and
South pole (right side panels) during the zonal GMF structure domination periods P$5_{21}$\,--\,P$5_{23}$ in cycles 21\,--\,23.}
   \label{zonal}
   \end{figure}

In Figures~\ref{lon}a and b, the polarity clusters of CHs and their association with the GMF
distribution and evolution are seen. To show the latitudinal distribution of CHs
and associated GMF during each period, the polar projection plots were created.
Polar projection distributions of CHs associated with the positive-polarity (red) and
negative-polarity (blue) photospheric magnetic fields and the GMF
are presented in Figures~\ref{cycle21}\,--\,\ref{zonal} seen
from the North pole (left side panels) and South pole (right side panels) for
different periods in cycles 21\,--\,23.

During the first periods (P$1_{21}$\,--\,P$1_{23}$), old-polarity polar
CHs still dominate at the poles,
but CHs associate with a new-polarity photospheric magnetic fields
are already at high latitudes and they are located at
preferred longitudes. The sectorial structure is not completely established.
The changes in GMF structure and CH distribution are rather chaotic
(Figures~\ref{lon}a, b, \ref{cycle21}a1, \ref{cycle21}a2\,--\,\ref{cycle23}a1, \ref{cycle23}a2).

During the periods P$2_{21}$\,--\,P$2_{23}$, P$3_{21}$\,--\,P$3_{23}$ the
new-polarity polar CHs are already completely formed.
Mid- and low-latitude CHs reveal a two-sectorial distribution.
The CH distribution coincides with that of the GMF.
Such a longitudinal distribution of the GMF and CHs indicates the increased
role of the axisymmetric GMF components at that time
The changes in the distribution of CHs associated with positive- and negative-polarity
magnetic fields coincides with the redistribution of the GMF
(Figures~\ref{gar}e, \ref{lon}a, b, and \ref{cycle21}b1, \ref{cycle21}b2, \ref{cycle21}c1, \ref{cycle21}c2\,--\,\ref{cycle23}b1, \ref{cycle23}b2, \ref{cycle23}c1, \ref{cycle23}c2).
During P3 periods, the two-sectorial structure is more pronounced in each cycle.

During the periods P$4_{21}$\,--\,P$4_{23}$, a four-sector structure is formed.
The sectorial and zonal harmonic spectral sum are nearly equal
(Figures~\ref{gar}a, b, \ref{lon}a, b and \ref{cycle21}d1, \ref{cycle21}d2\,--\,\ref{cycle23}d1, \ref{cycle23}d2).

During the periods P$5_{21}$\,--\,P$5_{23}$, the time of the zonal GMF structure domination,
the polarity of mid-latitude CHs matches that of the polar CHs in the corresponding
hemisphere. The polar and non-polar CHs are associated with the
GMF of the same polarity in each hemisphere (Figure~\ref{zonal}).
They reflect the zonal structure of the GMF with {\it m}=0, {\it n}=1.
The residuals of the sectorial structure can be seen in low-latitude GMF and CHs
(Figures~\ref{lon}a, b, and \ref{zonal}).
The large extensions are developed from the polar CHs to low latitudes at that time
(Figure~\ref{ext}). The life-time of some extensions exceeds several CRs.

\section{Coronal Hole Parameters and the Global Magnetic Field}  \label{secchparam}

\begin{figure}
   \centerline{\includegraphics[width=1.\textwidth,clip=]{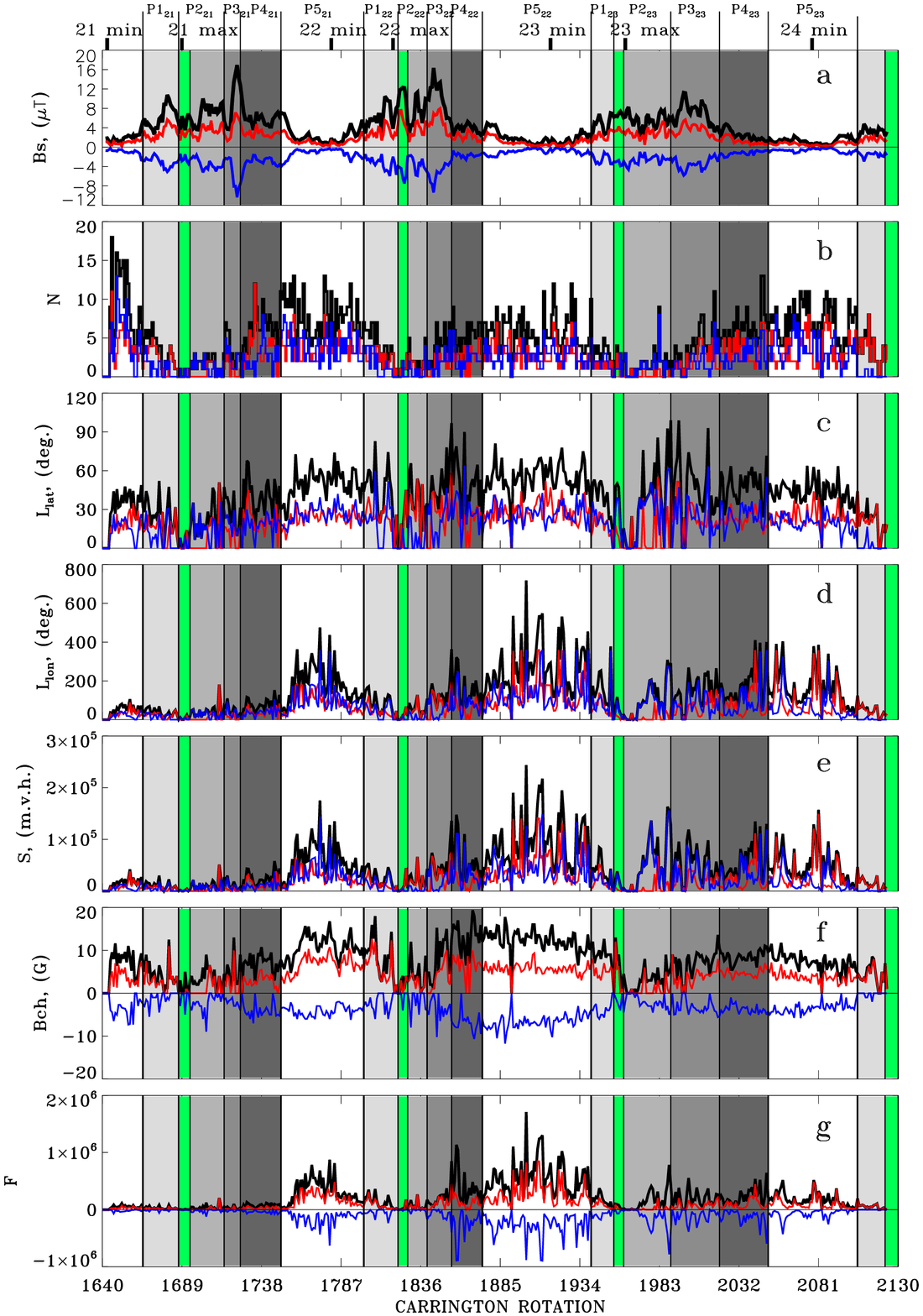}}
              \caption{(a) Positive-polarity (red) and negative-polarity (blue) CR-averaged GMF
              and their absolute value sum (black);
              (b) polar CH numbers;
              (c) polar CH latitudinal extension;
              (d) polar CH longitudinal extension;
              (e) polar CH area;
              (f) polar CH-associated photospheric magnetic field;
              (g) polar CH-associated magnetic flux.
             Red denotes the parameters of CHs associated with the positive-polarity magnetic fields.
             Blue denotes the parameters of CHs associated with the negative-polarity magnetic fields.
             Light-, mid-, and dark-grey  mark the P1\,--\,P4 periods of the sectorial GMF structure domination.
             Green indicates the magnetic-field polarity reversals at the North and South poles.}
   \label{chpol}
\end{figure}

\begin{figure}
   \centerline{\includegraphics[width=1.\textwidth,clip=]{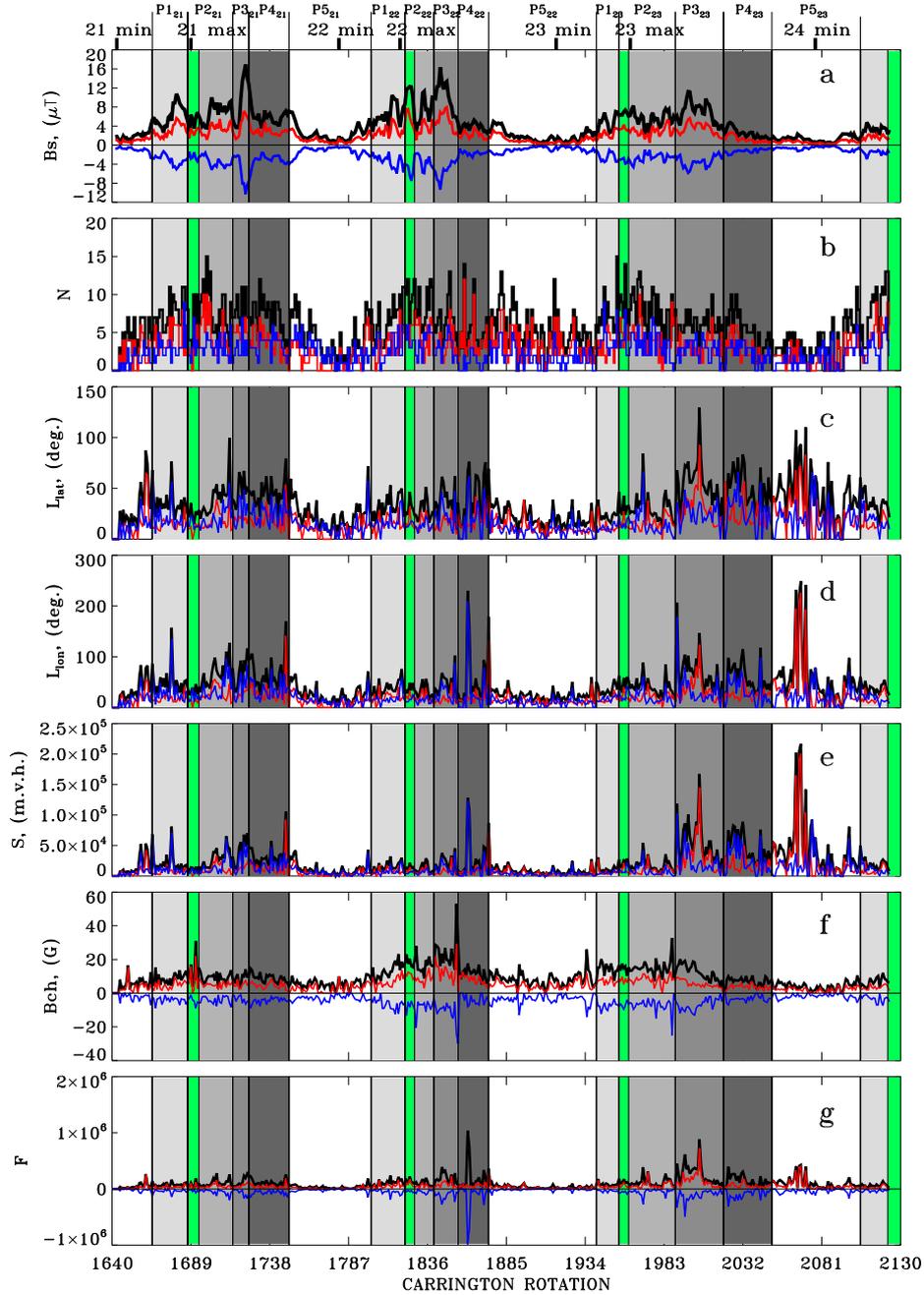}}
              \caption{(a) Positive-polarity (red) and negative-polarity (blue) CR-averaged GMF
              and their absolute value sum (black);
              (b) non-polar CH numbers;
              (c) non-polar CH latitudinal extension;
              (d) non-polar CH longitudinal extension;
              (e) non-polar CH area;
              (f) non-polar CH-associated photospheric magnetic field;
              (g) non-polar CH-associated magnetic flux.
             Red denotes the parameters of CHs associated with the positive-polarity magnetic fields.
             Blue denotes the parameters of CHs associated with the negative-polarity magnetic fields.
             Light-, mid-, and dark-grey mark the P1\,--\,P4 periods of the sectorial GMF structure domination.
             Green indicates the magnetic-field polarity reversals at the North and South poles.}
   \label{chnonpol}
\end{figure}

\begin{figure}
   \centerline{\includegraphics[width=1.\textwidth,clip=]{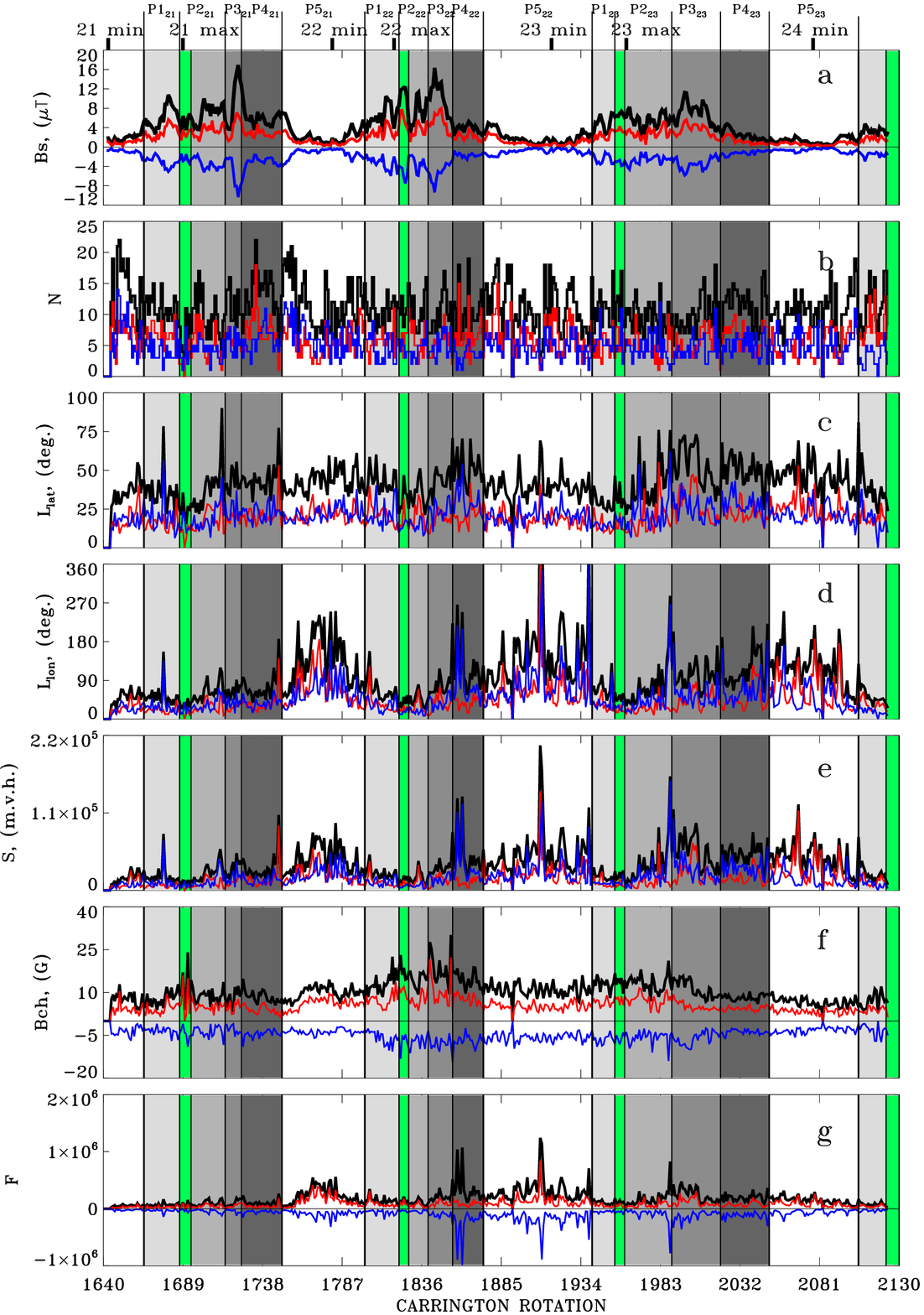}}
              \caption{(a) Positive-polarity (red) and negative-polarity (blue) CR-averaged GMF
              and their absolute value sum (black);
              (b) CH numbers;
              (c) CH latitudinal extension;
              (d) CH longitudinal extension;
              (e) CH area;
              (f) CH-associated photospheric magnetic field;
              (g) CH-associated magnetic flux.
              in (b) - (g) red denote the CHs associated with the positive-polarity magnetic fields and
            Red denotes the parameters of CHs associated with the positive-polarity magnetic fields.
            Blue denotes the parameters of CHs associated with the negative-polarity magnetic fields.
            Light-, mid-, and dark-grey mark the P1\,--\,P4 periods of the sectorial GMF structure domination.
            Green indicates the magnetic-field polarity reversals at the North and South poles.}
   \label{chall}
\end{figure}

As can be seen from Figure~\ref{chpol}, showing the polar CH parameters as well as
the positive- and negative-polarity and their absolute value sum of CR-averaged GMF evolution,
the number of polar CHs and their latitudinal and longitudinal extensions, areas,
and magnetic flux are maximal during the zonal GMF structure (P$5_{21}$\,--\,P$5_{23}$).
Non-polar GMF is minimal at that time.
There are additional peaks in polar CH latitudinal and longitudinal extensions,
areas, and magnetic flux during the end of period P$3_{22}$ in cycle 22, and
during periods P$2_{23}$\,--\,P$4_{23}$ in cycle 23.
Magnetic-field strength reaches a maximum in the period P$5_{21}$ (zonal GMF structure domination)
in cycle 21, and to the period P3 in cycles 22 and 23.
During polar magnetic-field reversals, the number of polar CHs is minimal
and all CH parameters are also at the lowest level (Figure~\ref{chpol}).
The polar magnetic-field sign changes are characterized by a decrease
in the GMF strength, and the magnitude of the sectorial components of
the GMF (Figure~\ref{gar}).

The number of non-polar CHs and all their parameters are maximal
during the sectorial GMF structure domination, and they are
at a low level at the time of the zonal GMF structure (Figure~\ref{chnonpol}).

The total CH number changes little during cycles 21\,--\,23 (Figure~\ref{chall}).
This is associated with a phase shift in the variation of the polar and non-polar CH numbers
by approximately half the cycle duration.
The polar CHs dominate in total CH longitudinal extension evolution and area.
The average total CH latitudinal extensions change rather chaotically.
The total CH flux and magnetic-field strength in CHs associated with the positive- and
negative-polarity photospheric magnetic fields is lower in cycle 23.
This is probably due to a decrease in the strength of
the polar magnetic field in cycle 23 (\opencite{Wang2009}; \opencite{Tlatov2014}).

\section{Conclusion} \label{secconclusion}

The comparison of CHs and GMF shows that
the GMF determines the CH evolution in cycles 21\,--\,23.

Comparison of the CH and GMF longitudinal distributions shows that
the CH cluster structure completely coincides with the GMF
longitudinal distribution. Positive-polarity CH locations coincide with positive-polarity GMF and
negative-polarity CH locations coincide with negative-polarity  GMF.
CH longitudinal distributions follow all configurations in the GMF.
Reorganizations of the GMF and CH cluster structures occur simultaneously
during a time interval of a few solar rotations.

CHs are divided into two groups: polar and non-polar CHs according to their
latitudinal locations. The number of polar CHs and their latitudinal and longitudinal extensions, areas,
and magnetic flux are maximal during the zonal GMF structure.
The number of non-polar CHs and all their parameters are maximal
during the sectorial GMF structure domination.
The total CH number changes little during cycles 21\,--\,23.
This is due to an approximately half a cycle phase shift in the polar
and non-polar CH numbers.

Two different type waves of non-polar coronal holes were revealed from their latitudinal distribution. The first one (waves 1) are short poleward waves. They trace the poleward motion of the unipolar photospheric magnetic
fields from approximately $35^{\circ}$ to the associated pole in each hemisphere
and the redevelopment of a new-polarity polar CH. Although they started the poleward movement
before the change of the polar magnetic field in the associated hemisphere,
they reached the pole after the polar reversal.

The other type are non-polar CH waves (waves 2) that form two sinusoidal branches. One branch is associated with the positive-polarity magnetic fields and the other with that of the negative-polarity. The complete period of the wave 2 CH waves was equal to $\approx$268 CRs ($\approx$22 years).
The  branches were in anti phase in each cycle.
Wave 2 CHs reached the highest latitudes and polar CH regions at the time
of the end of old-polarity magnetic-field domination at the associated pole, late at the declining phases.
The latitude-location cycle changes of the wave 2 CHs coincided with that of the
axisymmetric component of the solar dipole.
When two branches of positive- and negative-polarity magnetic fields, traced by wave 2 CHs,
went down below $\approx$$\pm35^{\circ}$ latitude, the sectorial structure of the GMF
was established. When both branches were above $\approx$$\pm35^{\circ}$ latitude,
the zonal structure was observed. The polarity of CH-associated magnetic
fields matched the polarity of the polar regions in the hemisphere that time.

The drifts in the GMF and CHs longitudinal distributions are the manifestation of
the slower rotation at the rising phases and faster rotation at the declining phases.
During the rising phases, old-polarity polar and low-latitude CHs, and
the associated photospheric magnetic fields,
were separated by the wave 1 CHs, that traced a new-polarity magnetic field.
At the late declining and minima phases,
the polarity of the polar and non-polar CHs, and the associated photospheric magnetic fields,
coincided. They created a common large-scale unipolar magnetic-field system.
Therefore, slow rotation rate of the polar regions affected the deceleration
of the rotation of mid- and low-latitude photospheric magnetic fields
and, consequently, the associated CHs.

%
 \begin{acks}

We acknowledged Dr. Tlatov A. G. and all Kislovodsk Mountain Astronomical Station
of Pulkovo Observatory team for the catalog of coronal holes used in this study.

Wilcox Solar Observatory data used in this study was obtained via
the web site
http://wso.\\stanford.edu at 2015:02:26 00:54:03 PST
courtesy of J.T. Hoeksema. The Wilcox Solar Observatory
is currently supported by NASA.

NSO/Kitt Peak data used here are produced cooperatively by
NSF/NOAO, NASA/GSFC, and NOAA/SEL.

The {\it Yohkoh} mission was developed and launched by ISAS/JAXA, Japan, with NASA and SERC/PPARC (UK) as international partners. This work made use of the {\it Yohkoh} Legacy data Archive at Montana State University, which is supported by NASA.

SOHO/EIT data were used. SOHO is a project of international cooperation between ESA and NASA.

\end{acks}

%
%
\bibliographystyle{spr-mp-sola}

 \bibliography{bilenko}

\begin{thebibliography}{51}
\ifx\bisbn     \undefined \def\bisbn  #1{ISBN #1}\fi
\ifx\binits    \undefined \def\binits#1{#1}\fi
\ifx\bauthor   \undefined \def\bauthor#1{#1}\fi
\ifx\batitle   \undefined \def\batitle#1{#1}\fi
\ifx\bjtitle   \undefined \def\bjtitle#1{\textit{#1}}\fi
\ifx\bvolume   \undefined \def\bvolume#1{\textbf{#1}}\fi
\ifx\byear     \undefined \def\byear#1{#1}\fi
\ifx\bissue    \undefined \def\bissue#1{#1}\fi
\ifx\bfpage    \undefined \def\bfpage#1{#1}\fi
\ifx\blpage    \undefined \def\blpage #1{#1}\fi
\ifx\burl      \undefined \def\burl#1{\textsf{#1}}\fi
\ifx\href      \undefined \def\href#1#2{\textsf{#2}}\fi
\ifx\betal     \undefined \def\betal{\textit{et al.}}\fi
\ifx\bctitle   \undefined \def\bctitle#1{#1}\fi
\ifx\beditor   \undefined \def\beditor#1{#1}\fi
\ifx\bbtitle   \undefined \def\bbtitle#1{\textit{#1}}\fi
\ifx\bedition  \undefined \def\bedition#1{#1}\fi
\ifx\bseriesno \undefined \def\bseriesno#1{\textbf{#1}}\fi
\ifx\blocation \undefined \def\blocation#1{#1}\fi
\ifx\bsertitle \undefined \def\bsertitle#1{\textit{#1}}\fi
\ifx\bsnm      \undefined \def\bsnm#1{#1}\fi
\ifx\bsuffix   \undefined \def\bsuffix#1{#1}\fi
\ifx\bparticle \undefined \def\bparticle#1{#1}\fi
\ifx\barticle  \undefined \def\barticle#1{}\fi
\ifx\binstitute  \undefined \def\binstitute#1{#1}\fi
\ifx\bpublisher  \undefined \def\bpublisher#1{#1}\fi
\ifx\doiurl    \undefined
  \def\doiurl#1{\href{http://dx.doi.org/#1}{\textsf{DOI}}}\fi
\ifx\arxivurl  \undefined
  \def\arxivurl#1{\href{http://arxiv.org/abs/#1}{\textsf{arXiv}}}\fi
\ifx\adsurl    \undefined
  \def\adsurl#1{\href{http://adsabs.harvard.edu/abs/#1}{\textsf{ADS}}}\fi
\ifx\botherref \undefined \def\botherref#1{}\fi
\ifx\url       \undefined \def\url#1{\textsf{#1}}\fi
\ifx\bchapter  \undefined \def\bchapter#1{}\fi
\ifx\bbook     \undefined \def\bbook#1{}\fi
\ifx\bcomment  \undefined \def\bcomment#1{#1}\fi
\ifx\oauthor   \undefined \def\oauthor#1{#1}\fi
\ifx\citeauthoryear \undefined\def \citeauthoryear#1{#1}\fi
\ifx\endbibitem\undefined \def\endbibitem{}\fi
\ifx\bconflocation  \undefined \def\bconflocation#1{#1} \fi

\bibitem[\protect\citeauthoryear{{Altschuler} and
  {Newkirk}}{1969}]{Altschuler1969}
\begin{barticle}
\bauthor{\bsnm{{Altschuler}}, \binits{M.D.}},
\bauthor{\bsnm{{Newkirk}}, \binits{G.}}:
\byear{1969},
\batitle{{Magnetic fields and the structure of the solar corona. I: methods of
  calculating coronal fields}}.
\bjtitle{\solphys}
\bvolume{9},
\bfpage{131}.
\doiurl{10.1007/BF00145734}.
\end{barticle}
\endbibitem

\bibitem[\protect\citeauthoryear{{Altschuler}
  \textit{et~al.}}{1975}]{Altschuler1975}
\begin{barticle}
\bauthor{\bsnm{{Altschuler}}, \binits{M.D.}},
\bauthor{\bsnm{{Trotter}}, \binits{D.E.}},
\bauthor{\bsnm{{Newkirk}}, \binits{G.J.}},
\bauthor{\bsnm{{Howard}}, \binits{R.}}:
\byear{1975},
\batitle{{Tabulation of the harmonic coefficients of the solar magnetic
  fields}}.
\bjtitle{\solphys}
\bvolume{41},
\bfpage{225}.
\doiurl{10.1007/BF00152968}.
\end{barticle}
\endbibitem

\bibitem[\protect\citeauthoryear{{Altschuler}
  \textit{et~al.}}{1977}]{Altschuler1977}
\begin{barticle}
\bauthor{\bsnm{{Altschuler}}, \binits{M.D.}},
\bauthor{\bsnm{{Levine}}, \binits{R.H.}},
\bauthor{\bsnm{{Stix}}, \binits{M.}},
\bauthor{\bsnm{{Harvey}}, \binits{J.}}:
\byear{1977},
\batitle{{High resolution mapping of the magnetic field of the solar corona}}.
\bjtitle{\solphys}
\bvolume{51},
\bfpage{345}.
\doiurl{10.1007/BF00216372}.
\end{barticle}
\endbibitem

\bibitem[\protect\citeauthoryear{{Belenko}}{2001}]{Belenko2001}
\begin{barticle}
\bauthor{\bsnm{{Belenko}}, \binits{I.A.}}:
\byear{2001},
\batitle{{Coronal hole evolution during 1996–1999}}.
\bjtitle{\solphys}
\bvolume{199},
\bfpage{23}.
\doiurl{10.1023/A:1010372926629}.
\end{barticle}
\endbibitem

\bibitem[\protect\citeauthoryear{{Bilenko}}{2002}]{Bilenko2002}
\begin{barticle}
\bauthor{\bsnm{{Bilenko}}, \binits{I.A.}}:
\byear{2002},
\batitle{{Coronal holes and the solar polar field reversal}}.
\bjtitle{\aap}
\bvolume{396},
\bfpage{657}.
\doiurl{10.1051/0004-6361:20021412}.
\end{barticle}
\endbibitem

\bibitem[\protect\citeauthoryear{{Bilenko}}{2004a}]{Bilenko2004a}
\begin{bchapter}
\bauthor{\bsnm{{Bilenko}}, \binits{I.A.}}:
\byear{2004}a,
\bctitle{{Formation and evolution of different type coronal holes}}.
In: \beditor{\bsnm{{Stepanov}}, \binits{A.V.}},
\beditor{\bsnm{{Benevolenskaya}}, \binits{E.E.}},
\beditor{\bsnm{{Kosovichev}}, \binits{A.G.}} (eds.)
\bbtitle{Multi-wavelength investigations of solar activity},
\bpublisher{Proceedings IAU Symposium No. 223}, \blocation{???},
\bfpage{373}.
\doiurl{10.1017/S1743921304006167}.
\end{bchapter}
\endbibitem

\bibitem[\protect\citeauthoryear{{Bilenko}}{2004b}]{Bilenko2004b}
\begin{barticle}
\bauthor{\bsnm{{Bilenko}}, \binits{I.A.}}:
\byear{2004}b,
\batitle{{Longitudinal distribution of coronal holes during 1976-2002}}.
\bjtitle{\solphys}
\bvolume{221},
\bfpage{261}.
\doiurl{10.1023/B:SOLA.0000035067.88819.40}.
\end{barticle}
\endbibitem

\bibitem[\protect\citeauthoryear{{Bilenko}}{2005}]{Bilenko2005}
\begin{barticle}
\bauthor{\bsnm{{Bilenko}}, \binits{I.A.}}:
\byear{2005},
\batitle{{Identification of the sources of the high-speed and low-speed streams
  of the solar wind}}.
\bjtitle{\ijga}
\bvolume{6},
\bfpage{GI1009}.
\doiurl{10.1029/2004GI000084}.
\end{barticle}
\endbibitem

\bibitem[\protect\citeauthoryear{{Bilenko}}{2012}]{Bilenko2012}
\begin{bchapter}
\bauthor{\bsnm{{Bilenko}}, \binits{I.A.}}:
\byear{2012},
\bctitle{{Statistical analysis of the structure and dynamics of coronal hole
  magnetic fields}}.
In: \beditor{\bsnm{{Ballester}}, \binits{P.}},
\beditor{\bsnm{{Egret}}, \binits{D.}},
\beditor{\bsnm{{Lorente}}, \binits{N.P.F.}} (eds.)
\bbtitle{Astronomical data analysis software and systems XXI},
\bpublisher{ASP, Vol. 461}, \blocation{???},
\bfpage{479}.
\end{bchapter}
\endbibitem

\bibitem[\protect\citeauthoryear{{Bilenko}}{2014}]{Bilenko2014}
\begin{barticle}
\bauthor{\bsnm{{Bilenko}}, \binits{I.A.}}:
\byear{2014},
\batitle{{Influence of the solar global magnetic-field structure evolution on
  CMEs}}.
\bjtitle{\solphys}
\bvolume{289},
\bfpage{4209}.
\doiurl{10.1007/s11207-014-0572-0}.
\end{barticle}
\endbibitem

\bibitem[\protect\citeauthoryear{{Bohlin} and {Sheeley}}{1978}]{Bohlin1978}
\begin{barticle}
\bauthor{\bsnm{{Bohlin}}, \binits{J.D.}},
\bauthor{\bsnm{{Sheeley}}, \binits{N.R.J.}}:
\byear{1978},
\batitle{{Extreme ultraviolet observations of coronal holes. II - Association
  of holes with solar magnetic fields and a model for their formation during
  the solar cycle}}.
\bjtitle{\solphys}
\bvolume{56},
\bfpage{125}.
\doiurl{10.1007/BF00152639}.
\end{barticle}
\endbibitem

\bibitem[\protect\citeauthoryear{{Bravo} and {Stewart}}{1997}]{Bravo1997}
\begin{barticle}
\bauthor{\bsnm{{Bravo}}, \binits{S.}},
\bauthor{\bsnm{{Stewart}}, \binits{G.A.}}:
\byear{1997},
\batitle{{Fast and slow wind from solar coronal holes}}.
\bjtitle{\apj}
\bvolume{489},
\bfpage{992}.
\end{barticle}
\endbibitem

\bibitem[\protect\citeauthoryear{{Bumba}, {Klva\u{n}a}, and
  {S\'ykora}}{1995}]{Bumba1995}
\begin{barticle}
\bauthor{\bsnm{{Bumba}}, \binits{V.}},
\bauthor{\bsnm{{Klva\u{n}a}}, \binits{M.}},
\bauthor{\bsnm{{S\'ykora}}, \binits{J.}}:
\byear{1995},
\batitle{{Coronal holes and their relation to the background and local magnetic
  fields}}.
\bjtitle{\aap}
\bvolume{298},
\bfpage{923}.
\end{barticle}
\endbibitem

\bibitem[\protect\citeauthoryear{{Chapman} and {Bartels}}{1940}]{Chapman1940}
\begin{bbook}
\bauthor{\bsnm{{Chapman}}, \binits{S.}},
\bauthor{\bsnm{{Bartels}}, \binits{J.}}:
\byear{1940},
\bbtitle{{Geomagnetism}},
\bpublisher{Geomagnetism, Oxford Univ. Press., 1940}, \blocation{???}.
\end{bbook}
\endbibitem

\bibitem[\protect\citeauthoryear{{Delaboudini\`{e}re}
  \textit{et~al.}}{1995}]{Delaboudiniere1995}
\begin{barticle}
\bauthor{\bsnm{{Delaboudini\`{e}re}}, \binits{J.-P.}},
\bauthor{\bsnm{{Artzner}}, \binits{G.E.}},
\bauthor{\bsnm{{Brunaud}}, \binits{J.}},
\bauthor{\bsnm{{Gabriel}}, \binits{A.H.}},
\bauthor{\bsnm{{Hochedez}}, \binits{J.F.}},
\bauthor{\bsnm{{Millier}}, \binits{F.}},
\bauthor{\bsnm{{Song}}, \binits{X.Y.}},
\bauthor{\bsnm{{Au}}, \binits{B.}},
\bauthor{\bsnm{{Dere}}, \binits{K.P.}},
\bauthor{\bsnm{{Howard}}, \binits{R.A.}},
\bauthor{\bsnm{{Kreplin}}, \binits{R.}},
\bauthor{\bsnm{{Michels}}, \binits{D.J.}},
\bauthor{\bsnm{{Moses}}, \binits{J.D.}},
\bauthor{\bsnm{{Defise}}, \binits{J.M.}},
\bauthor{\bsnm{{Jamar}}, \binits{C.}},
\bauthor{\bsnm{{Rochus}}, \binits{P.}},
\bauthor{\bsnm{{Chauvineau}}, \binits{J.P.}},
\bauthor{\bsnm{{Marioge}}, \binits{J.P.}},
\bauthor{\bsnm{{Catura}}, \binits{R.C.}},
\bauthor{\bsnm{{Lemen}}, \binits{J.R.}},
\bauthor{\bsnm{{Shing}}, \binits{L.}},
\bauthor{\bsnm{{Stern}}, \binits{R.A.}},
\bauthor{\bsnm{{Gurman}}, \binits{J.B.}},
\bauthor{\bsnm{{Neupert}}, \binits{W.M.}},
\bauthor{\bsnm{{Maucherat}}, \binits{A.}},
\bauthor{\bsnm{{Clette}}, \binits{F.}},
\bauthor{\bsnm{{Cugnon}}, \binits{P.}},
\bauthor{\bsnm{{van Dessel}}, \binits{E.L.}}:
\byear{1995},
\batitle{{EIT: extreme-ultraviolet imaging telescope for the SOHO mission}}.
\bjtitle{\solphys}
\bvolume{162},
\bfpage{291}.
\doiurl{10.1007/BF00733432}.
\end{barticle}
\endbibitem

\bibitem[\protect\citeauthoryear{{Dorotovi\u{c}}}{1996}]{Dorotovic1996}
\begin{barticle}
\bauthor{\bsnm{{Dorotovi\u{c}}}, \binits{I.}}:
\byear{1996},
\batitle{{Area of polar coronal holes and sunspot activity: years 1939-1993}}.
\bjtitle{\solphys}
\bvolume{167},
\bfpage{419}.
\doiurl{10.1007/BF00146350}.
\end{barticle}
\endbibitem

\bibitem[\protect\citeauthoryear{{Fox}, {McIntosh}, and
  {Wilson}}{1998}]{Fox1998}
\begin{barticle}
\bauthor{\bsnm{{Fox}}, \binits{P.}},
\bauthor{\bsnm{{McIntosh}}, \binits{P.}},
\bauthor{\bsnm{{Wilson}}, \binits{P.R.}}:
\byear{1998},
\batitle{{Coronal holes and the polar field reversals}}.
\bjtitle{\solphys}
\bvolume{177},
\bfpage{375}.
\doiurl{10.1023/A:1004939014025}.
\end{barticle}
\endbibitem

\bibitem[\protect\citeauthoryear{{Harvey} and {Recely}}{2002}]{Harvey2002}
\begin{barticle}
\bauthor{\bsnm{{Harvey}}, \binits{K.L.}},
\bauthor{\bsnm{{Recely}}, \binits{F.}}:
\byear{2002},
\batitle{{Polar coronal holes during cycles 22 and 23}}.
\bjtitle{\solphys}
\bvolume{211},
\bfpage{31}.
\doiurl{10.1023/A:1022469023581}.
\end{barticle}
\endbibitem

\bibitem[\protect\citeauthoryear{{Harvey}, {Sheeley}, and
  {Harvey}}{1982}]{Harvey1982}
\begin{barticle}
\bauthor{\bsnm{{Harvey}}, \binits{K.L.}},
\bauthor{\bsnm{{Sheeley}}, \binits{N.R.J.}},
\bauthor{\bsnm{{Harvey}}, \binits{J.W.}}:
\byear{1982},
\batitle{{Magnetic measurements of coronal holes during 1975-1980}}.
\bjtitle{\solphys}
\bvolume{79},
\bfpage{149}.
\doiurl{10.1007/BF00146979}.
\end{barticle}
\endbibitem

\bibitem[\protect\citeauthoryear{{Hess Webber}
  \textit{et~al.}}{2014}]{Hess2014}
\begin{barticle}
\bauthor{\bsnm{{Hess Webber}}, \binits{S.A.}},
\bauthor{\bsnm{{Karna}}, \binits{N.}},
\bauthor{\bsnm{{Pesnell}}, \binits{W.D.}},
\bauthor{\bsnm{{Kirk}}, \binits{M.S.}}:
\byear{2014},
\batitle{{Areas of polar coronal holes from 1996 through 2010}}.
\bjtitle{\solphys}
\bvolume{289},
\bfpage{4047}.
\doiurl{10.1007/s11207-014-0564-0}.
\end{barticle}
\endbibitem

\bibitem[\protect\citeauthoryear{{Hoeksema}}{}]{Hoeksema1984}
\begin{botherref}
\oauthor{\bsnm{{Hoeksema}}, \binits{J.T.}}:
\textit{{Structure and evolution of the large scale solar and heliospheric
  magnetic fields}},
Ph. D. Thesis, Stanford Univ., CA., 1984.
\end{botherref}
\endbibitem

\bibitem[\protect\citeauthoryear{{Hoeksema} and
  {Scherrer}}{1986}]{Hoeksema1986}
\begin{barticle}
\bauthor{\bsnm{{Hoeksema}}, \binits{J.T.}},
\bauthor{\bsnm{{Scherrer}}, \binits{P.H.}}:
\byear{1986},
\batitle{{An atlas of photospheric magnetic field observations and computed
  coronal magnetic fields: 1976-1985}}.
\bjtitle{\solphys}
\bvolume{105},
\bfpage{205}.
\doiurl{10.1007/BF00156388}.
\end{barticle}
\endbibitem

\bibitem[\protect\citeauthoryear{{Hoeksema} and
  {Scherrer}}{1988}]{Hoeksema1988}
\begin{barticle}
\bauthor{\bsnm{{Hoeksema}}, \binits{J.T.}},
\bauthor{\bsnm{{Scherrer}}, \binits{P.H.}}:
\byear{1988},
\batitle{{An atlas of photospheric magnetic field observations and computed
  coronal magnetic fields: 1976-1985}}.
\bjtitle{Solar - Geophysical Date (SGD)}
\bvolume{105},
\bfpage{no. 383}.
\end{barticle}
\endbibitem

\bibitem[\protect\citeauthoryear{{Ikhsanov} and {Ivanov}}{1999}]{Ikhsanov1999}
\begin{barticle}
\bauthor{\bsnm{{Ikhsanov}}, \binits{R.N.}},
\bauthor{\bsnm{{Ivanov}}, \binits{V.G.}}:
\byear{1999},
\batitle{{Properties of space and time distribution of solar coronal holes}}.
\bjtitle{\solphys}
\bvolume{188},
\bfpage{245}.
\doiurl{10.1023/A:1005109200233}.
\end{barticle}
\endbibitem

\bibitem[\protect\citeauthoryear{{Ikhsanov} and
  {Tavastsherna}}{2013}]{Ikhsanov2013}
\begin{barticle}
\bauthor{\bsnm{{Ikhsanov}}, \binits{R.N.}},
\bauthor{\bsnm{{Tavastsherna}}, \binits{K.S.}}:
\byear{2013},
\batitle{{High-latitude coronal holes and polar faculae in the 21st-23rd solar
  activity cycles}}.
\bjtitle{Geomagnetism and Aeronomy}
\bvolume{53},
\bfpage{896}.
\doiurl{10.1134/S0016793213070098}.
\end{barticle}
\endbibitem

\bibitem[\protect\citeauthoryear{{Ikhsanov} and
  {Tavastsherna}}{2015}]{Ikhsanov2015}
\begin{barticle}
\bauthor{\bsnm{{Ikhsanov}}, \binits{R.N.}},
\bauthor{\bsnm{{Tavastsherna}}, \binits{K.S.}}:
\byear{2015},
\batitle{{Latitude–temporal evolution of coronal holes in cycles 21-23}}.
\bjtitle{Geomagnetism and Aeronomy}
\bvolume{55},
\bfpage{877}.
\doiurl{10.1134/S0016793215070105}.
\end{barticle}
\endbibitem

\bibitem[\protect\citeauthoryear{{Insley}, {Moore}, and
  {Harrison}}{1995}]{Insley1995}
\begin{barticle}
\bauthor{\bsnm{{Insley}}, \binits{J.E.}},
\bauthor{\bsnm{{Moore}}, \binits{V.}},
\bauthor{\bsnm{{Harrison}}, \binits{R.A.}}:
\byear{1995},
\batitle{{The differential rotation of the corona as indicated by coronal
  holes}}.
\bjtitle{\solphys}
\bvolume{160},
\bfpage{1}.
\doiurl{10.1007/BF00679089}.
\end{barticle}
\endbibitem

\bibitem[\protect\citeauthoryear{{Ivanov} and {Obridko}}{2014}]{Ivanov2014}
\begin{barticle}
\bauthor{\bsnm{{Ivanov}}, \binits{E.V.}},
\bauthor{\bsnm{{Obridko}}, \binits{V.N.}}:
\byear{2014},
\batitle{{Role of the large-scale solar magnetic field structure in the global
  organization of solar activity}}.
\bjtitle{Geomagnetism and Aeronomy}
\bvolume{54},
\bfpage{996}.
\doiurl{10.1134/S0016793214080076}.
\end{barticle}
\endbibitem

\bibitem[\protect\citeauthoryear{{Levine}}{1977}]{Levine1977}
\begin{barticle}
\bauthor{\bsnm{{Levine}}, \binits{R.H.}}:
\byear{1977},
\batitle{{Evolution of photospheric magnetic field patterns during SKYLAB}}.
\bjtitle{\solphys}
\bvolume{54},
\bfpage{327}.
\doiurl{10.1007/BF00159923}.
\end{barticle}
\endbibitem

\bibitem[\protect\citeauthoryear{{McIntosh}, {Thompson}, and
  {Willock}}{1992}]{McIntosh1992}
\begin{barticle}
\bauthor{\bsnm{{McIntosh}}, \binits{P.S.}},
\bauthor{\bsnm{{Thompson}}, \binits{R.J.}},
\bauthor{\bsnm{{Willock}}, \binits{E.C.}}:
\byear{1992},
\batitle{{A 600-day periodicity in solar coronal holes}}.
\bjtitle{\nat}
\bvolume{360},
\bfpage{322}.
\doiurl{10.1038/360322a0}.
\end{barticle}
\endbibitem

\bibitem[\protect\citeauthoryear{{Miralles}, {Cranmer}, and
  {Kohl}}{2001}]{Miralles2001}
\begin{barticle}
\bauthor{\bsnm{{Miralles}}, \binits{M.P.}},
\bauthor{\bsnm{{Cranmer}}, \binits{S.R.}},
\bauthor{\bsnm{{Kohl}}, \binits{J.L.}}:
\byear{2001},
\batitle{{Ultraviolet coronagraph spectrometer observations of a high-latitude
  coronal hole with high oxygen temperatures and the next solar cycle
  polarity}}.
\bjtitle{\apj}
\bvolume{560},
\bfpage{L193}.
\doiurl{10.1086/324314}.
\end{barticle}
\endbibitem

\bibitem[\protect\citeauthoryear{{Miralles}, {Cranmer}, and
  {Kohl}}{2002}]{Miralles2002}
\begin{bchapter}
\bauthor{\bsnm{{Miralles}}, \binits{M.P.}},
\bauthor{\bsnm{{Cranmer}}, \binits{S.R.}},
\bauthor{\bsnm{{Kohl}}, \binits{J.L.}}:
\byear{2002},
\bctitle{{Cyclic variation in the plasma properties of coronal holes}}.
In: \beditor{\bsnm{{Wilso}}, \binits{A.}} (ed.)
\bbtitle{SOHO 11. From solar min to max: half a solar cycle with SOHO},
\bpublisher{ESA SP-508}, \blocation{???},
\bfpage{351}.
\end{bchapter}
\endbibitem

\bibitem[\protect\citeauthoryear{{Miralles}, {Cranmer}, and
  {Kohl}}{2006}]{Miralles2006}
\begin{bchapter}
\bauthor{\bsnm{{Miralles}}, \binits{M.P.}},
\bauthor{\bsnm{{Cranmer}}, \binits{S.R.}},
\bauthor{\bsnm{{Kohl}}, \binits{J.L.}}:
\byear{2006},
\bctitle{{Coronal hole properties during the first decade of UVCS/SOHO}}.
In: \beditor{\bsnm{{Lacoste}}, \binits{H.}} (ed.)
\bbtitle{SOHO 17 - 10 years of SOHO and beyond},
\bpublisher{ESA SP-617, ESTEC, The Netherlands}, \blocation{???},
\bfpage{15}.
\end{bchapter}
\endbibitem

\bibitem[\protect\citeauthoryear{{Nolte} \textit{et~al.}}{1976}]{Nolte1976}
\begin{barticle}
\bauthor{\bsnm{{Nolte}}, \binits{J.T.}},
\bauthor{\bsnm{{Krieger}}, \binits{A.S.}},
\bauthor{\bsnm{{Timothy}}, \binits{A.F.}},
\bauthor{\bsnm{{Gold}}, \binits{R.E.}},
\bauthor{\bsnm{{Roelof}}, \binits{E.C.}},
\bauthor{\bsnm{{Vaiana}}, \binits{G.}},
\bauthor{\bsnm{{Lazarus}}, \binits{A.J.}},
\bauthor{\bsnm{{Sullivan}}, \binits{J.D.}},
\bauthor{\bsnm{{McIntosh}}, \binits{P.S.}}:
\byear{1976},
\batitle{{Coronal holes as sources of solar wind}}.
\bjtitle{\solphys}
\bvolume{46},
\bfpage{303}.
\doiurl{10.1007/BF00149859}.
\end{barticle}
\endbibitem

\bibitem[\protect\citeauthoryear{{Obridko} and {Shelting}}{1989}]{Obridko1989}
\begin{barticle}
\bauthor{\bsnm{{Obridko}}, \binits{V.N.}},
\bauthor{\bsnm{{Shelting}}, \binits{B.D.}}:
\byear{1989},
\batitle{{Coronal holes as indicators of large-scale magnetic fields in the
  corona}}.
\bjtitle{\solphys}
\bvolume{124},
\bfpage{73}.
\doiurl{10.1007/BF00146520}.
\end{barticle}
\endbibitem

\bibitem[\protect\citeauthoryear{{Obridko}, {Shelting}, and
  {Livshits}}{2011}]{Obridko2011}
\begin{barticle}
\bauthor{\bsnm{{Obridko}}, \binits{V.N.}},
\bauthor{\bsnm{{Shelting}}, \binits{B.D.}},
\bauthor{\bsnm{{Livshits}}, \binits{I.M.}}:
\byear{2011},
\batitle{{Relationship between the parameters of coronal holes and hihg-speed
  solar wind streams over an activity cycle}}.
\bjtitle{\solphys}
\bvolume{270},
\bfpage{297}.
\doiurl{10.1007/s11207-011-9753-2}.
\end{barticle}
\endbibitem

\bibitem[\protect\citeauthoryear{{Sanchez-Ibarra} and
  {Barraza-Paredes}}{1992}]{Sanchez1992}
\begin{bbook}
\bauthor{\bsnm{{Sanchez-Ibarra}}, \binits{A.}},
\bauthor{\bsnm{{Barraza-Paredes}}, \binits{M.}}:
\byear{1992},
\bbtitle{{Catalog of coronal holes, 1970-1991, report UAG-102}},
\bpublisher{Boulder: World Data Center A for solar-terrestrial physics,
  National Geophysical Data Center, 1992}, \blocation{???}.
\end{bbook}
\endbibitem

\bibitem[\protect\citeauthoryear{{Schatten}, {Wilcox}, and
  {Ness}}{1969}]{Schatten1969}
\begin{barticle}
\bauthor{\bsnm{{Schatten}}, \binits{K.H.}},
\bauthor{\bsnm{{Wilcox}}, \binits{J.M.}},
\bauthor{\bsnm{{Ness}}, \binits{N.F.}}:
\byear{1969},
\batitle{{A model of interplanetary and coronal magnetic fields}}.
\bjtitle{\solphys}
\bvolume{6},
\bfpage{442}.
\doiurl{10.1007/BF00146478}.
\end{barticle}
\endbibitem

\bibitem[\protect\citeauthoryear{{Stix}}{1977}]{Stix1977}
\begin{barticle}
\bauthor{\bsnm{{Stix}}, \binits{M.}}:
\byear{1977},
\batitle{{Coronal holes and the large-scale solar magnetic field}}.
\bjtitle{\aap}
\bvolume{59},
\bfpage{73}.
\end{barticle}
\endbibitem

\bibitem[\protect\citeauthoryear{{Tavastsherna} and
  {Polyakow}}{2014}]{Tavastsherna2014}
\begin{barticle}
\bauthor{\bsnm{{Tavastsherna}}, \binits{K.S.}},
\bauthor{\bsnm{{Polyakow}}, \binits{E.V.}}:
\byear{2014},
\batitle{{Coronal holes, large-scale magnetic field, and activity complexes in
  solar cycle 23}}.
\bjtitle{Geomagnetism and Aeronomy}
\bvolume{54},
\bfpage{953}.
\doiurl{10.1134/S0016793214070135}.
\end{barticle}
\endbibitem

\bibitem[\protect\citeauthoryear{{Tavastsherna} and
  {Tlatov}}{2004}]{Tavastsherna2004}
\begin{bchapter}
\bauthor{\bsnm{{Tavastsherna}}, \binits{K.S.}},
\bauthor{\bsnm{{Tlatov}}, \binits{A.G.}}:
\byear{2004},
\bctitle{{Properties of the magnetic field in the coronal holes in solar cycle
  23}}.
In: \beditor{\bsnm{{Stepanov}}, \binits{A.V.}},
\beditor{\bsnm{{Benevolenskaya}}, \binits{E.E.}},
\beditor{\bsnm{{Kosovichev}}, \binits{A.G.}} (eds.)
\bbtitle{Multi-wavelength investigations of solar activity},
\bpublisher{IAU Symposium No. 223}, \blocation{???},
\bfpage{301}.
\doiurl{10.1017/S1743921304006039}.
\end{bchapter}
\endbibitem

\bibitem[\protect\citeauthoryear{{Temmer}, {Vr\u{s}nak}, and
  {Veronig}}{2007}]{Temmer2007}
\begin{barticle}
\bauthor{\bsnm{{Temmer}}, \binits{M.}},
\bauthor{\bsnm{{Vr\u{s}nak}}, \binits{B.}},
\bauthor{\bsnm{{Veronig}}, \binits{A.M.}}:
\byear{2007},
\batitle{{Periodic appearance of coronal holes and the related variation of
  solar wind parameters}}.
\bjtitle{\solphys}
\bvolume{241},
\bfpage{371}.
\doiurl{10.1007/s11207-007-0336-1}.
\end{barticle}
\endbibitem

\bibitem[\protect\citeauthoryear{{Timothy}, {Krieger}, and
  {Vaiana}}{1975}]{Timothy1975}
\begin{barticle}
\bauthor{\bsnm{{Timothy}}, \binits{A.F.}},
\bauthor{\bsnm{{Krieger}}, \binits{A.S.}},
\bauthor{\bsnm{{Vaiana}}, \binits{G.S.}}:
\byear{1975},
\batitle{{The structure and evolution of coronal holes}}.
\bjtitle{\solphys}
\bvolume{42},
\bfpage{135}.
\doiurl{10.1007/BF00153291}.
\end{barticle}
\endbibitem

\bibitem[\protect\citeauthoryear{{Tlatov}, {Tavastsherna}, and
  {Vasil'eva}}{2014}]{Tlatov2014}
\begin{barticle}
\bauthor{\bsnm{{Tlatov}}, \binits{A.}},
\bauthor{\bsnm{{Tavastsherna}}, \binits{K.}},
\bauthor{\bsnm{{Vasil'eva}}, \binits{V.}}:
\byear{2014},
\batitle{{Coronal holes in solar cycles 21 to 23}}.
\bjtitle{\solphys}
\bvolume{289},
\bfpage{1349}.
\doiurl{10.1007/s11207-013-0387-4}.
\end{barticle}
\endbibitem

\bibitem[\protect\citeauthoryear{{Tsuneta} \textit{et~al.}}{1991}]{Tsuneta1991}
\begin{barticle}
\bauthor{\bsnm{{Tsuneta}}, \binits{S.}},
\bauthor{\bsnm{{Acton}}, \binits{L.}},
\bauthor{\bsnm{{Bruner}}, \binits{M.}},
\bauthor{\bsnm{{Lemen}}, \binits{J.}},
\bauthor{\bsnm{{Brown}}, \binits{W.}},
\bauthor{\bsnm{{Caravalho}}, \binits{R.}},
\bauthor{\bsnm{{Catura}}, \binits{R.}},
\bauthor{\bsnm{{Freeland}}, \binits{S.}},
\bauthor{\bsnm{{Jurcevich}}, \binits{B.}},
\bauthor{\bsnm{{Morrison}}, \binits{M.}},
\bauthor{\bsnm{{Ogawara}}, \binits{Y.}},
\bauthor{\bsnm{{Hirayama}}, \binits{T.}},
\bauthor{\bsnm{{Owens}}, \binits{J.}}:
\byear{1991},
\batitle{{The soft X-ray telescope for the SOLAR-A mission}}.
\bjtitle{\solphys}
\bvolume{136},
\bfpage{37}.
\doiurl{10.1007/BF00151694}.
\end{barticle}
\endbibitem

\bibitem[\protect\citeauthoryear{{Varsik}, {Wilson}, and
  {Li}}{1999}]{Varsik1999}
\begin{barticle}
\bauthor{\bsnm{{Varsik}}, \binits{J.R.}},
\bauthor{\bsnm{{Wilson}}, \binits{P.R.}},
\bauthor{\bsnm{{Li}}, \binits{Y.}}:
\byear{1999},
\batitle{{High-resolution studies of the solar polar magnetic fields}}.
\bjtitle{\solphys}
\bvolume{184},
\bfpage{223}.
\doiurl{10.1023/A:1005179027771}.
\end{barticle}
\endbibitem

\bibitem[\protect\citeauthoryear{{Waldmeier}}{1981}]{Waldmeier1981}
\begin{barticle}
\bauthor{\bsnm{{Waldmeier}}, \binits{M.}}:
\byear{1981},
\batitle{{Cyclic variations of the polar coronal hole}}.
\bjtitle{\solphys}
\bvolume{70},
\bfpage{251}.
\doiurl{10.1007/BF00151332}.
\end{barticle}
\endbibitem

\bibitem[\protect\citeauthoryear{{Wang} and {Sheeley}}{1990}]{Wang1990}
\begin{barticle}
\bauthor{\bsnm{{Wang}}, \binits{Y.M.}},
\bauthor{\bsnm{{Sheeley}}, \binits{N.R.J.}}:
\byear{1990},
\batitle{{Magnetic flux transport and the sunspot-cycle evolution of coronal
  holes and their wind streams}}.
\bjtitle{\apj}
\bvolume{365},
\bfpage{372}.
\doiurl{10.1086/169492}.
\end{barticle}
\endbibitem

\bibitem[\protect\citeauthoryear{{Wang}, {Robbrecht}, and
  {Sheeley}}{2009}]{Wang2009}
\begin{barticle}
\bauthor{\bsnm{{Wang}}, \binits{Y.-M.}},
\bauthor{\bsnm{{Robbrecht}}, \binits{E.}},
\bauthor{\bsnm{{Sheeley}}, \binits{N.R.J.}}:
\byear{2009},
\batitle{{On the weakening of the polar magnetic fields during solar cycle
  23}}.
\bjtitle{\apj}
\bvolume{707},
\bfpage{1372–1386}.
\doiurl{10.1088/0004-637X/707/2/1372}.
\end{barticle}
\endbibitem

\bibitem[\protect\citeauthoryear{{Webb}, {Davis}, and
  {McIntosh}}{1984}]{Webb1984}
\begin{barticle}
\bauthor{\bsnm{{Webb}}, \binits{D.F.}},
\bauthor{\bsnm{{Davis}}, \binits{J.M.}},
\bauthor{\bsnm{{McIntosh}}, \binits{P.S.}}:
\byear{1984},
\batitle{{Observations of the reappearance of polar coronal holes and the
  reversal of the polar magnetic field}}.
\bjtitle{\solphys}
\bvolume{92},
\bfpage{109}.
\doiurl{10.1007/BF00157239}.
\end{barticle}
\endbibitem

\bibitem[\protect\citeauthoryear{{Zhao}, {Hoeksema}, and
  {Scherrer}}{1999}]{Zhao1999}
\begin{barticle}
\bauthor{\bsnm{{Zhao}}, \binits{X.P.}},
\bauthor{\bsnm{{Hoeksema}}, \binits{J.T.}},
\bauthor{\bsnm{{Scherrer}}, \binits{P.H.}}:
\byear{1999},
\batitle{{Changes of the boot-shaped coronal hole boundary during whole sun
  month near sunspot minimum}}.
\bjtitle{\jgr}
\bvolume{104},
\bfpage{9735}.
\doiurl{10.1029/1998JA900010}.
\end{barticle}
\endbibitem

\end{thebibliography}
%
%
%
%

\end{article}
\end{document}